\documentclass[aps,prb,twocolumn,showpacs,epsfig,floats]{revtex4}
\usepackage{amsmath}
\usepackage{color}
\usepackage{graphicx}
\usepackage{amsfonts}
\usepackage{mathrsfs}
\usepackage{amsmath}
\usepackage{hyperref}
\usepackage{pifont}

\begin{document}
\title{Floquet topological transition by unpolarized light}

\author{Bhaskar Mukherjee$^{1}$}

\affiliation { $^{1}$Theoretical Physics Department, Indian
Association for the Cultivation of Science, Jadavpur,
Kolkata-700032, India.}

\date{\today}
\begin{abstract}

We study Floquet topological transition in irradiated graphene when the polarization of incident light changes randomly with time. 
We numerically confirm that the noise averaged time evolution operator approaches a steady value in the limit of exact
Trotter decomposition of the whole period where incident light has different polarization at each interval of the decomposition. 
This steady limit is found to coincide with time-evolution
operator calculated from the noise-averaged Hamiltonian. We observe that at the six corners (Dirac($K$) point) of the hexagonal Brillouin zone of graphene
random Gaussian noise strongly modifies the phaseband structure
induced by circularly polarized light whereas in zone-center ($\Gamma$ point) even a strong noise isn't able to do
the same. This can be understood by analyzing the deterministic noise averaged Hamiltonian which has a different Fourier structure
as well as lesser no of symmetries compared to the noise-free one. In 1D systems noise is found to renormalize the drive amplitude only.

\end{abstract}

\maketitle
\section{Introduction}\label{I}

Realizing topological phenomenon in solid state system has been one of the major topic in condensed matter physics since
the discovery of IQHE in 2D semiconductor devices\cite{iqhe1}. These materials are model system for 2D non-interacting
electron gas which under the application of strong magnetic field forms highly gapped Landau levels at low temperature. 
This results in very precise quantization of Hall conductance\cite{iqhe2,iqhe3} and supports robust conducting chiral states at the edges\cite{iqhe4,iqhe5}. Later it was shown
that the magnetic field is not necessary and one can also observe such phenomenon
in systems described by tight-binding Hamiltonians\cite{haldane}. The so called \textquotedblleft Haldane Model\textquotedblright
describe electrons hopping in a honeycomb lattice threaded by periodic magnetic flux with zero net flux. The resulting complex
hopping is difficult to implement experimentally and it is only recently that the advancement in ultra-cold atomic systems have made such
experiments possible\cite{expt1}.
To avoid such complicated implementation of the Haldane model and thus realize Chern insulating states more easily, a possible
alternative way, namely \textquotedblleft irradiation of electromagnetic wave on graphene\textquotedblright, is proposed recently to achieve the essential 
goal of time reversal symmetry breaking. 

Graphene is a gapless 2D Dirac system which can open up a gap at the Dirac Point under irradiation of circularly polarized light\cite{jap1,jap2}.
This resulting new state, termed as Floquet topological insulator was found later in many other systems\cite{gil,roderich}. It is also
detectable by various transport signatures \cite{takashi,arijit,gil2}.
These are steady states of periodically driven non-equilibrium systems\cite{rigol1,rigol2,das1,das2} which recently gained
tremendous attention because of it's potential to create new phases. These phases can hardly be found in their equilibrium
counterparts. Traditional bulk-boundary correspondence was extended to Floquet topological systems taking into account the periodicity of the Floquet spectrum\cite{rudner1,rudner2}. Experimental verification of such
states has already been achieved using both time and angle resolved photoemission spectroscopy(PES)\cite{gedik,gavensky} and also in photonic systems\cite{rechtsman,sebabrata}. 

Throughout the last decade a large number of studies of real time dynamics in closed quantum systems have extended the notions of universality from equilibrium to 
non-equilibrium via Kibble-Zurek scaling\cite{ksengupta}. Further studies show that the qualitative nature of these scalings can be completely
reversed by introducing noise in the drive\cite{anirban}. In these studies the Heisenberg equation of motion picks up a dephasing term due to averaging
over different noise realizations which leads to non-unitary dynamics. Recently in equilibrium systems it has been shown that periodicity in space (i.e the crystal structure) is not necessary to get
topological behavior and one can also see it in amorphous systems\cite{adhip}. Analogously one can ask at this point that what would happen
in Floquet systems if time periodicity of the Hamiltonian is broken due to the presence of noise in the drive. Several studies in this direction in models decomposable in free fermions
have already revealed that the nature of the asymptotic steady state depends on the type of aperiodic protocol\cite{sourav}. Further 
some analytical studies show that disorder-averaging can be avoided for a special class of protocols\cite{utso}. 

Influenced by this kind of works we plan to study the fate of the Floquet topological systems when the smooth time variation of 
incident electromagnetic wave is broken by the insertion of a random phase in one of the component of vector potential. This kind of noise is 
always there in a typical experiment if the setup to produce polarized light isn't calibrated properly. Moreover
such noise can also be generated artificially using synthetic gauge fields.
We term this kind of monochromatic wave as unpolarized light in the sense that the associated Lissajous figures keeps on changing with time.
The central results of this work can be summarized as follows. We show that depending on the spatial dimension of the problem
Floquet topological transitions can be influenced by the random change in polarization of incident light. For 
graphene we find that the transitions at Dirac(K) point are significantly modified compared to $\Gamma$ point. The origin
of this effect can be understood to be due to a fundamental change in Fourier structure of the noise-averaged time-dependent 
Hamiltonian at K point. At low frequencies of the incident radiation, it is well known that symmetries of the underlying 
Hamiltonian is crucial for topological transition\cite{us}. In the presence of noise, we find such symmetries to be broken. 
Interestingly, in contrast to standard expectation, we find that few of these symmetries are restored in the noise-averaged
Hamiltonian. This symmetry restoration has impact on the self-averaging limit in this parameter regime.
Finally for a 1D model($p$-wave superconducting wire), using a non-trivial drive protocol, we show that even a strong noise (large 
standard deviation) can't prohibit the transition.

The rest of the paper is planned as follows. In Sec.\ref{II} we introduce our protocol for irradiated graphene and plot the 
results(phasebands) for numerical disorder averaging. 
In Sec.\ref{IIA} we establish the existence of self-averaging limit which suggests
numerical averaging is meaningful and can be mimicked by the ensemble averaged Hamiltonian. This is followed by possible 
explanation of the deviation from noise free (circularly polarized case) behavior separately in high and low frequency 
regime in Sec.\ref{IIB} and Sec.\ref{IIC} respectively. Next, in Sec.\ref{III}, we shows results for 1D systems.
Finally we conclude and discuss possible experimental scenarios in Sec.\ref{IV}.
\section{Irradiated Graphene}\label{II}
 We consider graphene irradiated by electromagnetic wave defined by the vector potential $\bold{A}=A_0(\cos(\omega t+\phi(t)),
 \sin(\omega t))$. One have to further assume it to be space independent in graphene plane to keep the integrability of the problem intact. The 
$\phi=0$(circularly polarized) case is well studied in the literature\cite{arijit2}. We allow $\phi$ to be a normally distributed 
random variable with mean $\mu$ and standard deviation $\sigma$ at each instant of time which gives rise to its unpolarized nature. If one wish to produce this vector potential in lab
then this kind of noise will be inherently present as random experimental error. The normalized probability distribution 
of $\phi$ at each time instant t is given by
\begin{equation}\label{(1)}
 P(\phi)=\frac{1}{\sqrt{2\pi}\sigma}e^{-\frac{(\phi-\mu)^2}{2\pi \sigma^2}}
\end{equation}
$\mu$ can be any real number within the interval 
($-\pi \le \mu \le \pi$). Here we will concentrate on the special value $\mu=0$ (i.e this is the value of $\mu$ in all plot). This will allow us to directly compare the result with circularly polarized case. 

The time-dependent graphene Hamiltonian(for each k-mode) after Peierls's substitution with this protocol becomes
\begin{equation}H(\bold{k},t)=\left(\begin{array}{cc}
0 & Z(\bold{k},t)\\
Z^*(\bold{k},t) & 0
\end{array}\right)\nonumber
\end{equation}
where $Z(\bold{k},t)=-\gamma (2e^{i\frac{\tilde{k_x}}{2}}\cos(\frac{\sqrt{3}\tilde{k_y}}{2})+e^{-i\tilde{k_x}})$ 
and $\bold{\tilde{k}}=\bold{k}+e\bold{A}$

Next we calculate the time-evolution operator over one time period($T$) for each k-mode by dividing the period in N parts
\begin{eqnarray}\label{(2)}
 U_k(T,0)&=&T_te^{-i\int_0^TH_k(t')dt'}\nonumber\\
&=&e^{-iH_k(T-\delta t)\delta t}e^{-iH_k(T-2\delta t)\delta t}.....e^{-iH_k(2\delta t)\delta t}\nonumber\\
&&e^{-iH_k(\delta t)\delta t}
\end{eqnarray}
where $T_t$ denotes time-ordered product and $\delta t=T/N$ is a very small but fixed time interval. Such decomposition introduces Trotter error which gets reduced with increasing N
and reproduces the exact U for the chosen continuous drive in the $N\rightarrow \infty$ limit. We calculate the time-dependent 
Hamiltonian at each partition by drawing $\phi$ from a normal distribution and using Eq.\ref{(2)} get $U(T,0)$ for one particular noise realization.
We then average over several such realizations numerically and get the noise averaged time evolution operator
\begin{equation}\label{(3)}
 \langle U_k(T,0)\rangle=\langle T_te^{-i\int_0^TH_k(t')dt'} \rangle.
\end{equation}
Eq.\ref{(3)} has a self-averaging limit\cite{Lobejko}, in the sense that all four elements of $\langle U(T,0) \rangle$ 
goes to some steady value with increasing
no of partitions (N). We shall discuss this in more details in the next sub-section. 

In Fig.1 we plot the phasebands ($\Phi(T)$)
obtained using $\cos(\Phi(T))=Re[\langle U(T) \rangle_{11}]$.
\begin{figure}
 \includegraphics[width=0.49\linewidth]{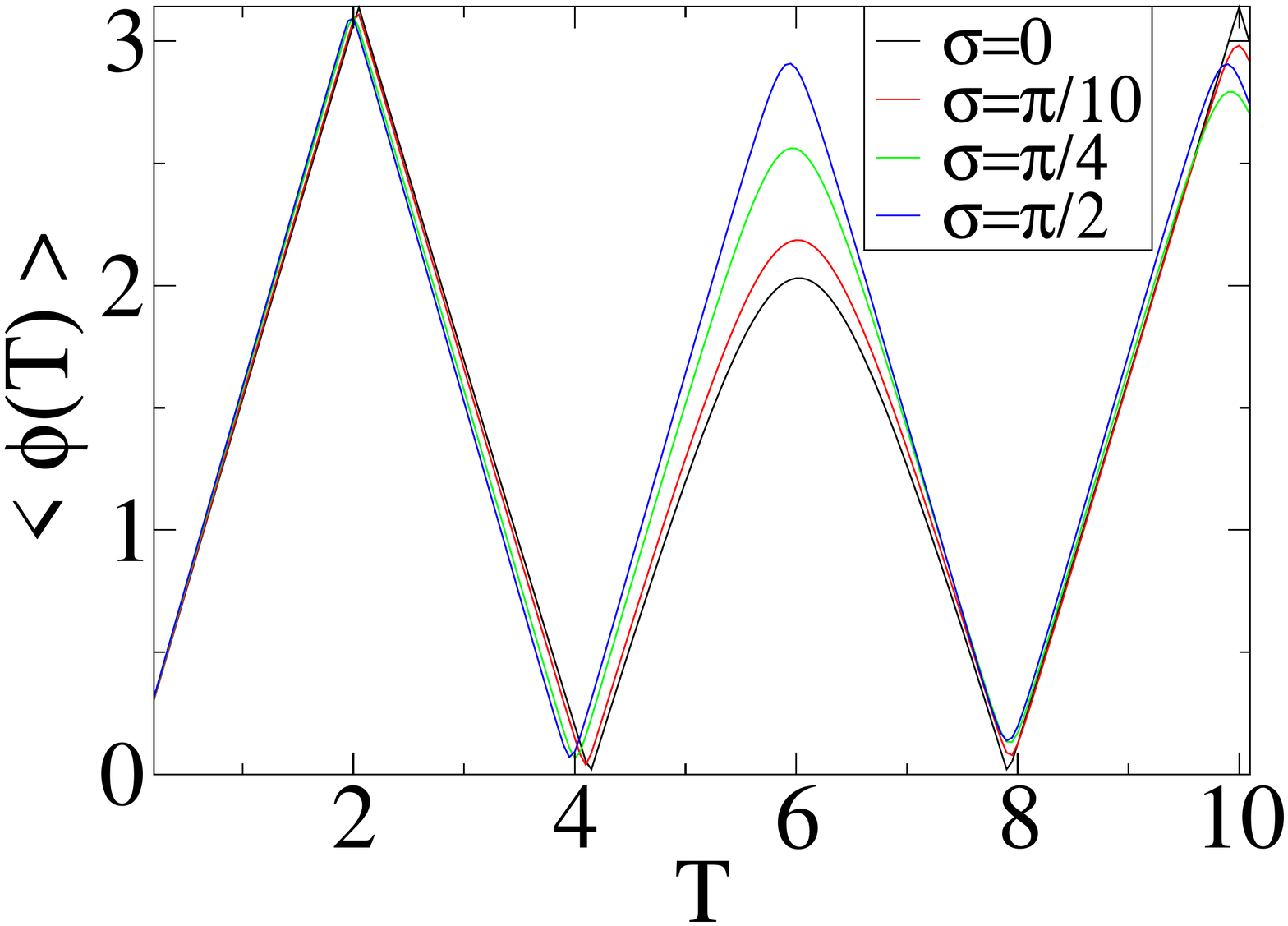}
\includegraphics[width=0.49\linewidth]{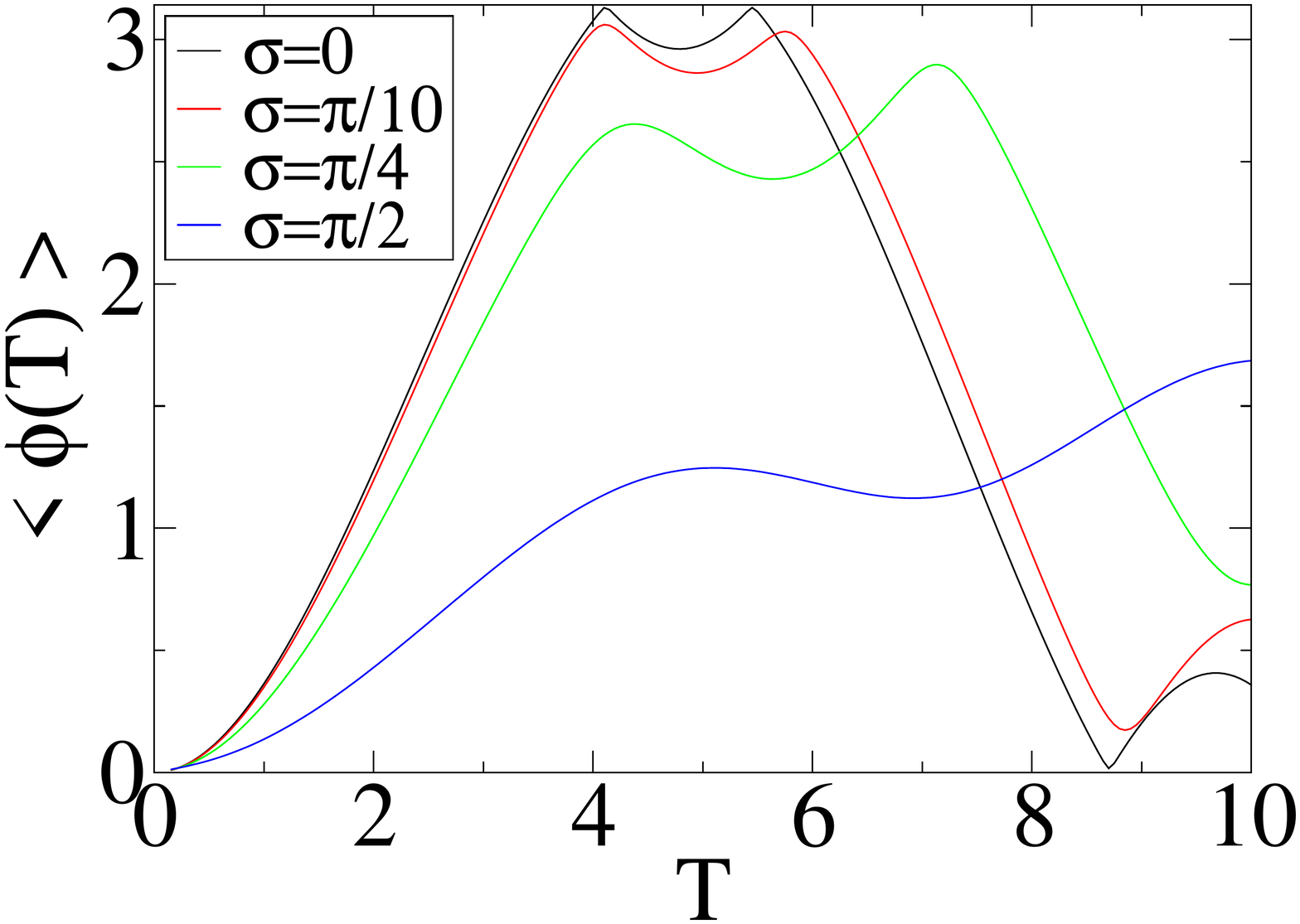}
\caption{Noise averaged phasebands vs T for $\Gamma$ point(left) (at $\alpha=1.5$) and for $K$ point(right) (at $\alpha=2.0$) for various 
values of standard deviation($\sigma$). N=1000, no of sample=1000 and $\alpha=eA_0/c$}
\end{figure}
One can see with increasing magnitude of random noise
the phasebands gets modified but we recover the results for pure circularly polarized light in $\sigma\rightarrow0$ limit as expected. 
We find that the phasebands remain almost unchanged for $\Gamma$ point for a broad range of parameter
values; however at $K$ point, they are strongly modified by the noise. We calculate
Chern number of the lower Floquet band using the eigenfunctions of $\langle U(T) \rangle$ in a discretized Brillouin zone.
The plot is shown in Fig.2. We find that the transitions (position of integer jump in Chern number) can sustain an appreciable amount
of temporal noise and merely gets shifted in parameter space but very strong noise (large $\sigma$) abolish them.
\begin{figure}
  \includegraphics[width=0.49\linewidth]{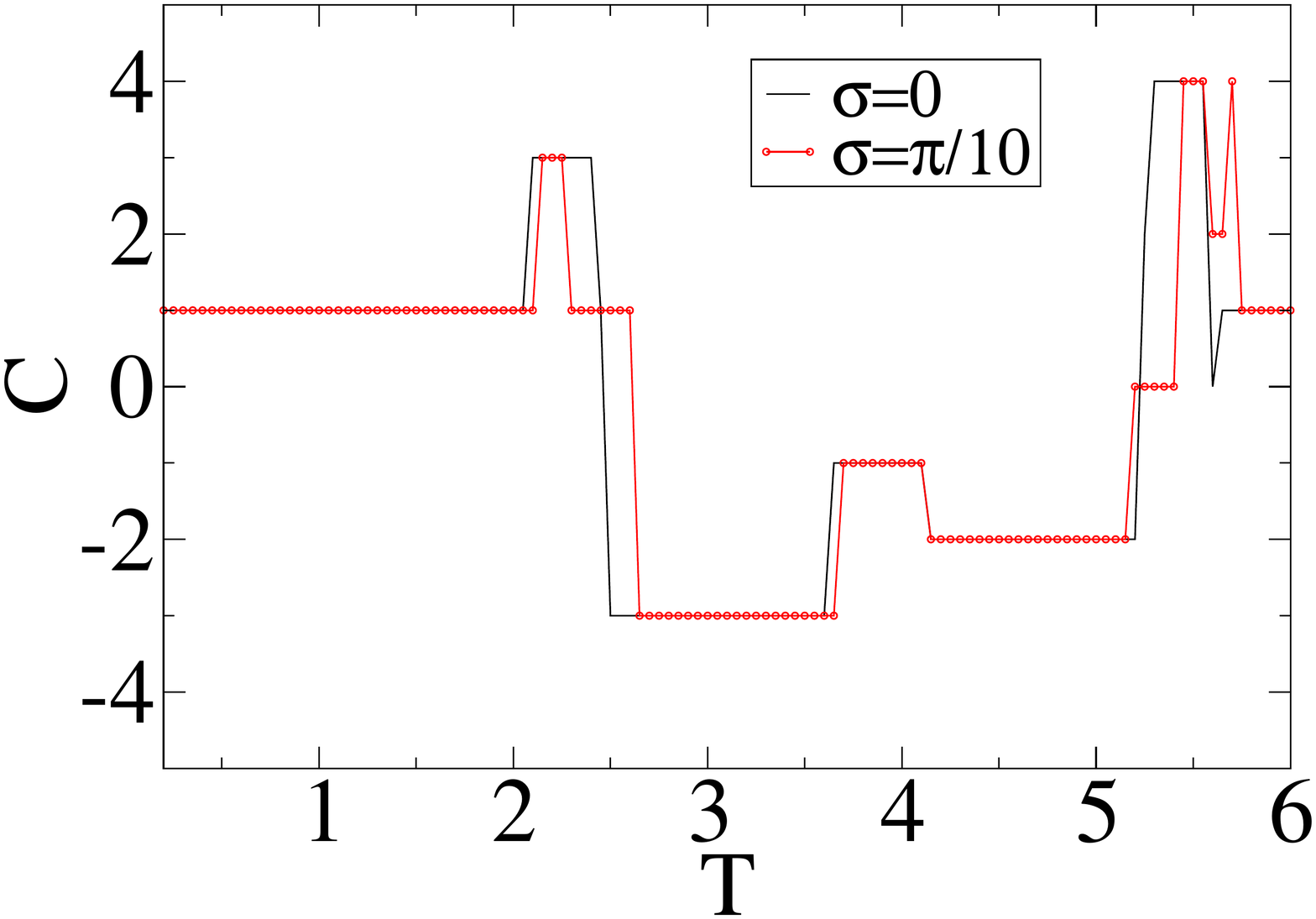}
\includegraphics[width=0.49\linewidth]{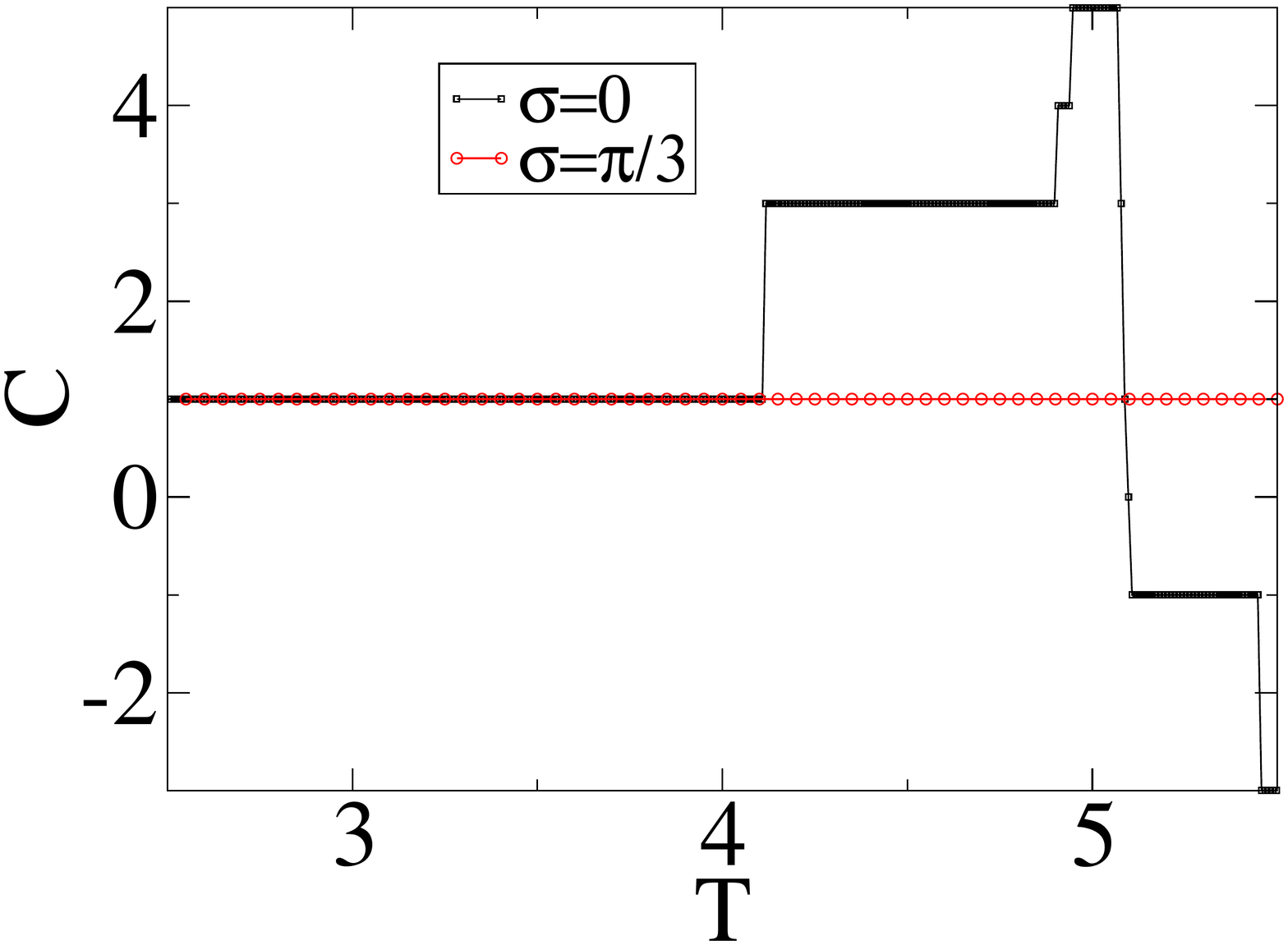}
\caption{Chern number of the noise averaged lower Floquet band for $\Gamma$ (left) and $K$ (right) point. Others parameters are same as 
in Fig.1}
\end{figure}

\subsection{Ensemble averaged Hamiltonian}\label{IIA}
In this subsection we explore the possibility of constructing a deterministic Hamiltonian 
such that time-evolution operator constructed using it resembles the noise averaged time-evolution operator. In a recent work\cite{Lobejko} Lobejko
{\it et al} have showed rigorously that the difference of ensemble averaged time-evolution operator and the time-evolution operator constructed 
by the  ensemble averaged Hamiltonian scales as $O(\frac{1}{N})$ for a certain class of protocols. For these protocols the ensemble averaged
Hamiltonian at two different time commutes which they have termed as \textquotedblleft commutation in statistical sense\textquotedblright. They further extends
the applicability of above theorem to some simple non-commuting Hamiltonian by numerical simulations. But unlike those cases
irradiated graphene contains the noise term within the argument of complicated trigonometric functions. Hence the ensemble averaged Hamiltonian 
can not be obtained here simply by substituting $\phi$ by it's mean value. Therefore we explicitly
calculate the ensemble-averaged Hamiltonian for irradiated graphene at time t
\begin{equation}\label{(4)}
 \langle H_k(t) \rangle=\int_{-\infty}^{\infty}P(\phi)H_k(\phi,t)d\phi
\end{equation}
 with $P(\phi)$ in Eq.\ref{(1)} we get using Jacobi-Anger relations\cite{google}.
\begin{widetext}
\begin{eqnarray}\label{(5)}
 \langle Z(\bold{k},t) \rangle&=&-\gamma (2e^{i\frac{k_x}{2}}\cos(\frac{\sqrt{3}(k_y+\alpha \sin(\omega t))}{2})[J_0(\frac{\alpha}{2})
+2\sum_{n=1}^{\infty}i^nJ_n(\frac{\alpha}{2})e^{-\frac{n^2\sigma^2}{2}}\cos(n(\omega t+\mu))]+e^{-ik_x}[J_0(\alpha)+\nonumber\\
&&2\sum_{n=1}^{\infty}(-i)^nJ_n(\alpha)e^{-\frac{n^2\sigma^2}{2}}\cos(n(\omega t+\mu))])
\end{eqnarray}
\end{widetext}
Using this we numerically calculate the Frobenius norm of the distance between $\langle U(H(t))\rangle$ and $U(\langle H(t)\rangle)$
\begin{equation}\label{(6)}
 D_N=\parallel\langle Te^{-i\int_0^TH(t')dt'}\rangle-Te^{-i\int_0^T\langle H(t') \rangle dt'}\parallel
\end{equation}
and the same norm for the corresponding variance matrix
\begin{equation}\label{(7)}
 S_N=\parallel\langle(Te^{-i\int_0^TH(t')dt'}-Te^{-i\int_0^T\langle H(t') \rangle dt'})^2\rangle\parallel
\end{equation}
where N is the no of partitions used to calculate (using Eq.\ref{(2)} and \ref{(3)}) each quantities inside the norm. These are two appropriate quantities to measure
the deviation of the time-evolution operator in different noise realizations. We see power law fall of both $D_N$ and $S_N$ in no of partitions(N)(see Fig.3) which suggest self-averaging limit exists here. 
It is only in this limit that the disorder averaging is meaningful in dynamical systems. This is in close analogy to
equilibrium disordered systems where for each disorder realization some amount of deviation (from the mean) is introduced in all physical observable
due to the finite size of the system but these deviations get canceled when averaged out over several disorder realizations and thus 
helps to achieve the thermodynamic result fast. Here in dynamical system finite no of partition(N) play the role of finite system size
and the thermodynamic limit corresponds to the continuous drive ($N \rightarrow \infty$). Vanishing of $S_N$ in large N also implies
the equivalence
\begin{equation}\label{(8)}
 \cos(\langle\Phi(T)\rangle)\equiv\langle\cos(\Phi(T))\rangle
\end{equation}
which we have used throughout the paper. In Fig.3 note that $D_N$ and $S_N$ have larger values
at $K$ point compared to $\Gamma$ point for small N. This is related to the fact that time dependent Hamiltonian of
irradiated graphene at $K$ point is more complicated than at $\Gamma$ point due to the presence of lesser no of 
symmetries\cite{us}. Larger the complexity larger N one need to use to reduce these errors.
\begin{figure}
{\includegraphics[width=0.45 \columnwidth]{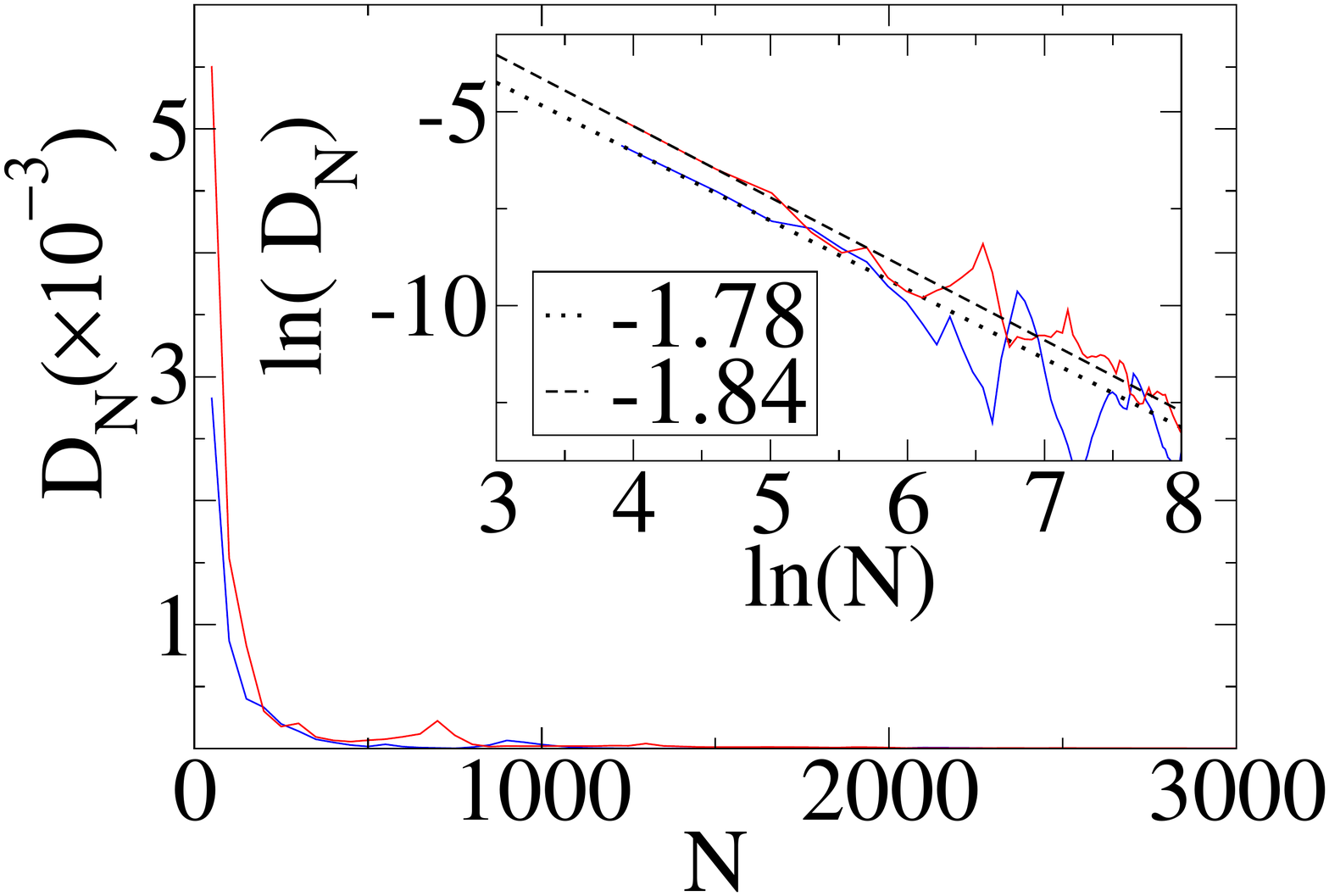}}
{\includegraphics[width=0.45 \columnwidth]{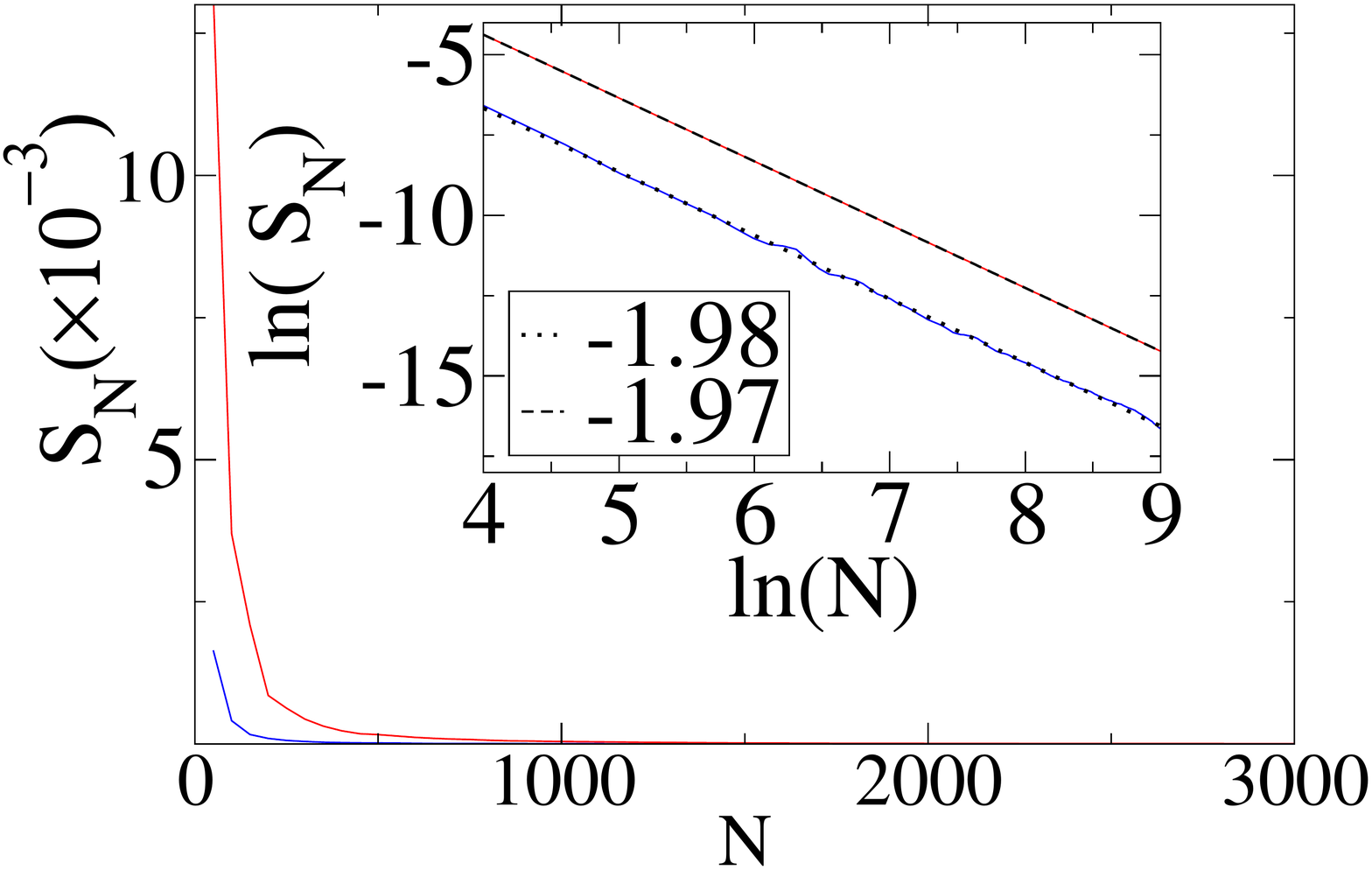}} \\
{\includegraphics[width=0.45 \columnwidth]{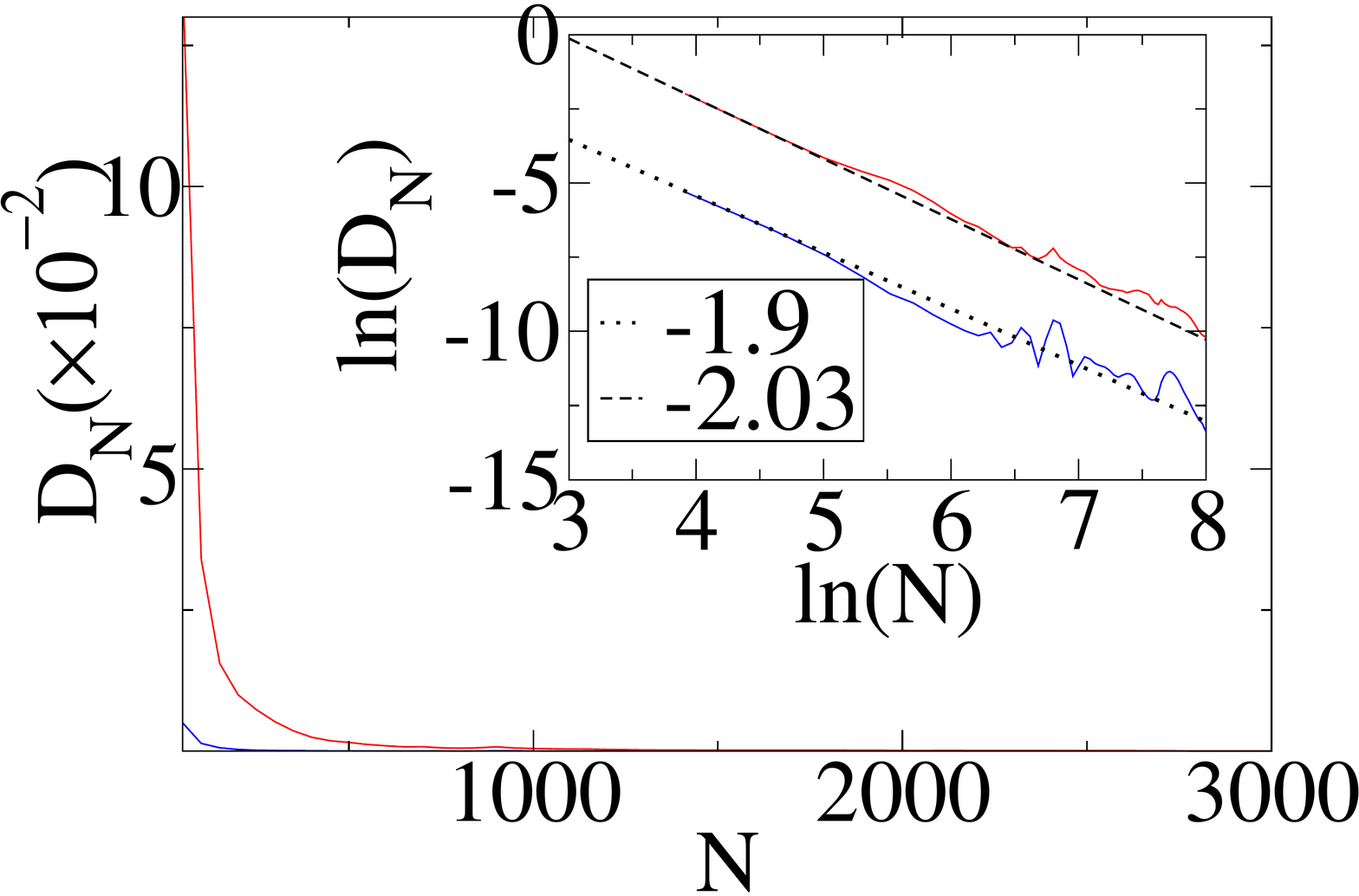}}
{\includegraphics[width=0.45 \columnwidth]{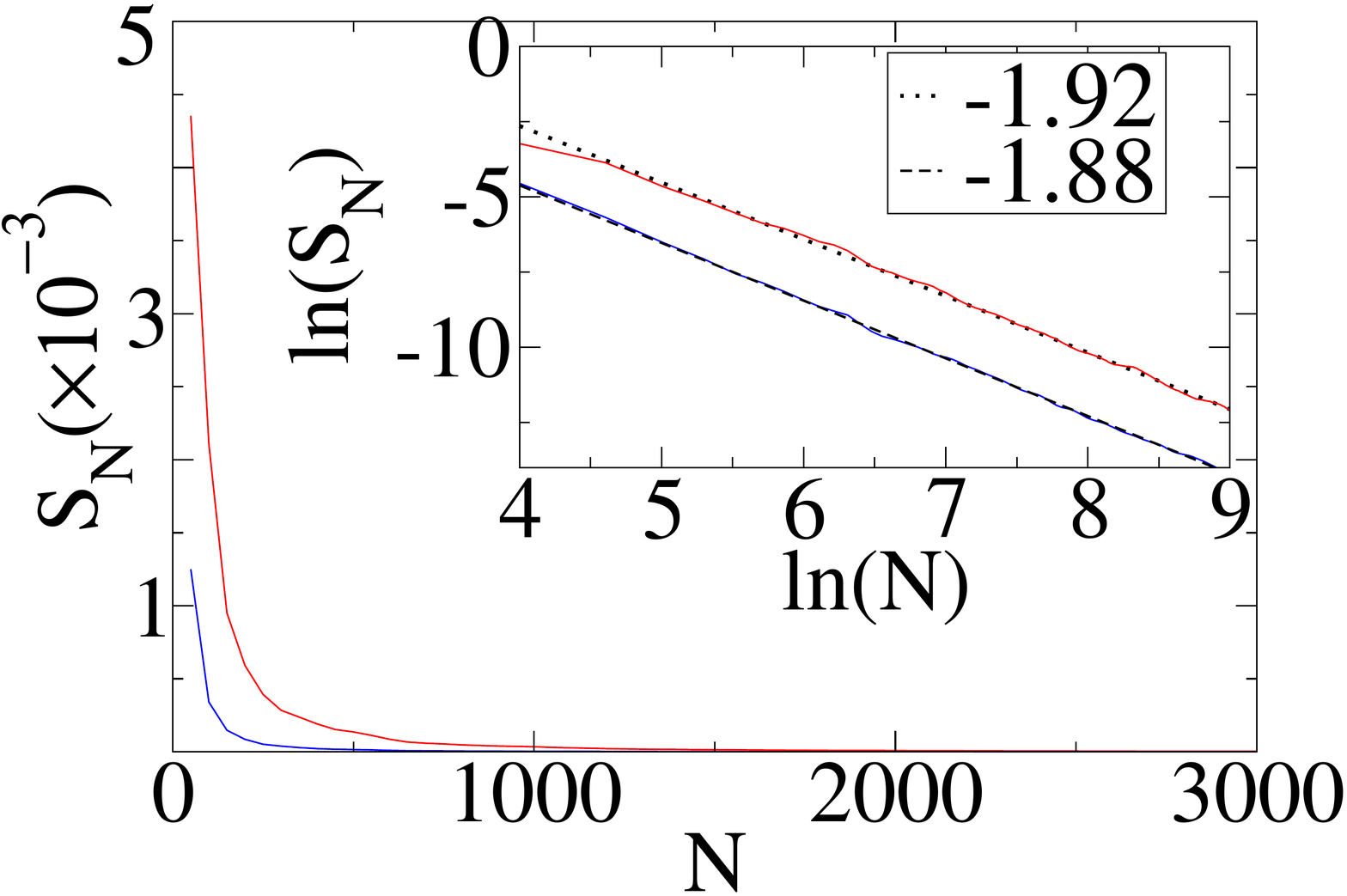}} \\
\caption{Fall of $D_N$(upper left panel) and $S_N$(upper right panel) with N for $\Gamma$ point at $\alpha=1.5$ and $T=4.0$. Same for the 
Dirac point in lower left and right panel at $\alpha=2$ and $T=4.0$. No of sample=1000 and $\sigma$ for blue and red curve is $\pi/10$ and $\pi/3$ 
respectively for all 4 panel. Slope of the linear fit is mentioned in the insets.}
\end{figure}
This power law fall suggests that the time consuming numerical disorder averaging can be avoided by the use of ensemble averaged Hamiltonian to calculate $U(T,0)$
with a sufficiently large no of partitions of whole period. We further demonstrate this by explicitly comparing the phasebands from
both this way in Fig.4.
\begin{figure}
  \includegraphics[width=0.49\linewidth]{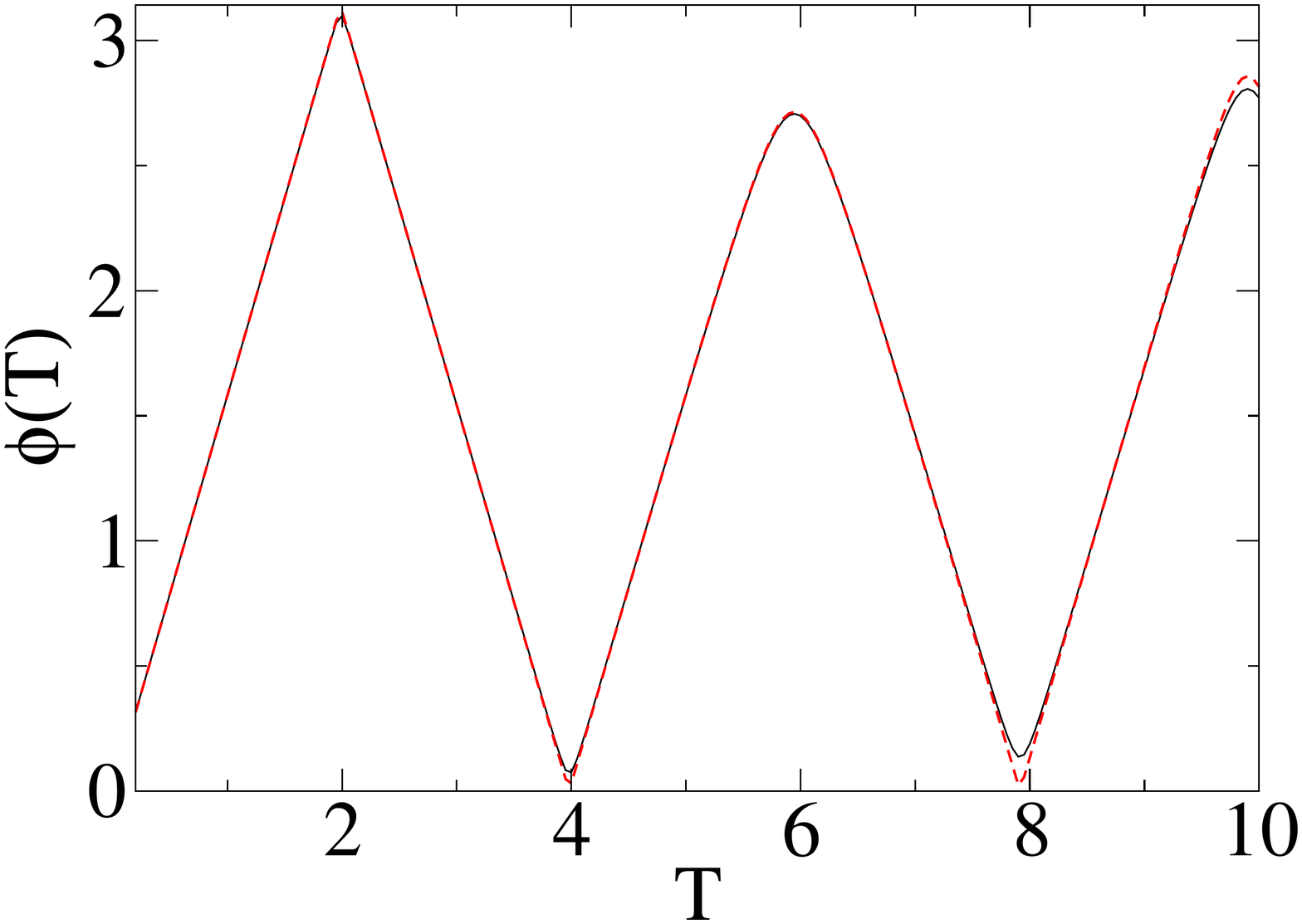}
\includegraphics[width=0.49\linewidth]{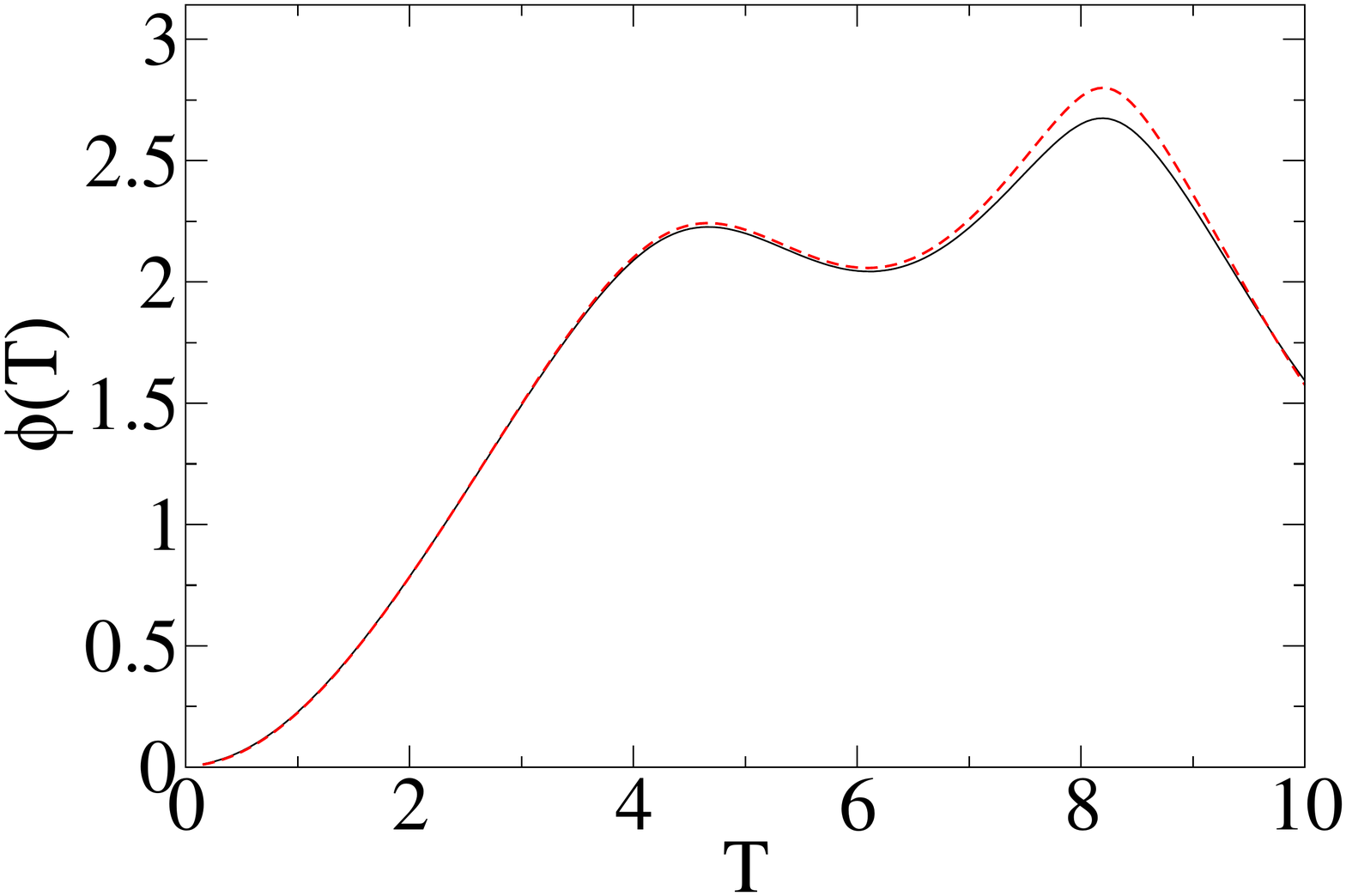} 
\caption{Comparison of phasebands obtained by numerically disorder averaging of $U(T,0)$ operator (black-solid line) and by using
the ensemble averaged $H(t)$ to calculate $U(T,0)$ operator (red-dashed line) for $\Gamma$ point (left panel) and for $K$ point (right panel).
Relevant parameters are same as in Fig.3.}
\end{figure}
Our next target is to understand better why in some cases a weak noise is sufficient to abolish all the transition (as in $K$ point) where
as in some other cases(as in $\Gamma$ point) even a strong noise just causes a shift of the crossing positions and nothing more 
than that. We will do it by analyzing the ensemble averaged Hamiltonian (Eq.\ref{(5)}) in two different frequency regime.

\subsection{High frequency, Floquet formalism}\label{IIB}
The Floquet formalism allows one to treat a periodic time-dependent problem as a 
time-independent eigenvalue problem. The cost of this is to deal with an infinite dimensional Hilbert space which is a vector space 
of T periodic functions and also known as Sambe space. The representation of Floquet Hamiltonian (related to $U(T,0)$ by $U(T,0)=e^{-iH_FT}$)
in this basis is defined by the following matrix elements
\begin{equation}\label{(9)}
 H_{i,j}^{m,n}=m\omega \delta_{mn} \delta_{ij}+\frac{1}{T}\int_0^Te^{-i(m-n)\omega t'}H_{ij}(t')dt'
\end{equation}
where $(m,n)$ is row and column index of different square blocks each of size $(H_1\times H_1)$ where $H_1$ is the Hilbert space dimension 
of the equilibrium problem (2 for each k-mode in our case) and $(i,j)$ denotes position of each matrix element within one such block. For 
numerical purposes one can truncate this matrix after some order which depends on details of the problem especially the absolute 
value of maximum order of the Fourier components (of time-dependent Hamiltonian) with non-vanishing coefficient. One also needs to
increase the truncation dimension with decreasing frequency. Following this prescription one can safely truncate the Floquet Hamiltonian
in zero-th order at $\Gamma$ point(where one has a $2\times2$ $H_F$) and in 1st order at $K$ point( where one has a $6\times6$ $H_F$) for
high frequencies and low Amplitude of radiation\cite{jap1,arijit2}. Thus one gets expressions of Floquet conduction band 
($\Phi(T)$) in 1st quasi-energy BZ for the noise free (circularly polarized) case with hopping-amplitude($\gamma$) set to unity
\begin{equation}\label{(10)}
\hskip -2.7cm \Phi(\Gamma,T)=3J_0(\alpha)T
\end{equation}
\begin{equation}\label{(11)}
 \Phi(K,T)=\frac{\sqrt{4\pi^2+36J_1^2(\alpha)T^2}-2\pi}{2}
\end{equation}
Next we aim to calculate some simplified expression of phaseband for the unpolarized light using the ensemble averaged Hamiltonian in
some suitable parameter regime. We can sufficiently simplify Eq.\ref{(5)} for strong noise. Note that though $\phi$ appears
as argument of trigonometric functions due to it's random nature at each instant of time $\phi[\mu,\sigma]$ and $\phi[\mu+2n\pi,\sigma+2p\pi]$
will not give same time evolution operator. Using $e^{-\frac{n^2\sigma ^2}{2}}\approx0$ for large $\sigma$ in 
Eq.\ref{(5)} we get 
\begin{eqnarray}\label{(12)}
 \langle Z(\bold{k},t)\rangle\mid_{\sigma\gg0}&\approx&-\gamma(2J_0(\frac{\alpha}{2})e^{ikx}\cos(\frac{\sqrt{3}}{2}(ky+\nonumber\\
&&\alpha \sin(\omega t)))+J_0(\alpha)e^{-ikx})
\end{eqnarray}
for $\Gamma$ point this gives a Hamiltonian proportional to $\sigma_x$ only and hence one simply gets the phaseband 
\begin{equation}\label{(13)}
 \Phi(\Gamma,T)=\int_0^T\langle Z(\Gamma,t')\rangle dt'
\end{equation}
the integrand is difficult but again using Jacobi-Anger relations we get(taking $\gamma=1$)
\begin{eqnarray}\label{(14)}
 \Phi(\Gamma,T)&=&(2J_0(\frac{\alpha}{2})J_0(\frac{\sqrt{3}\alpha}{2})+J_0(\alpha))T+\nonumber\\
 &&4J_0(\frac{\alpha}{2})\sum_{n=1}^{\infty}J_{2n}(\frac{\sqrt{3}\alpha}{2})\int_0^T\cos(2n\omega t')dt'\nonumber\\
 &=&(2J_0(\frac{\alpha}{2})J_0(\frac{\sqrt{3}\alpha}{2})+J_0(\alpha))T
\end{eqnarray}
similarly for $K$ point we get
\begin{equation}\label{(15)}
 \Phi(K,T)=(J_0(\alpha)-J_0(\frac{\alpha}{2})J_0(\frac{\sqrt{3}\alpha}{2}))T
\end{equation}
we compare cosines of Floquet bands for circularly polarized($\sigma=0$) and unpolarized($\sigma\gg0$) case in Fig.5. The 
functional behavior of these two bands do not change much for $\Gamma$ point whereas for $K$ point they show drastically 
different behavior. This huge change for $K$ point is due to the fact that strong noise (highly unpolarized light) changes the 
lowest non-vanishing Fourier component of $\langle H_K(t) \rangle$ from 1 to 0 and thus reduces the 
effective Sambe space dimension from 6 to 2. These changes make the Floquet band at $K$ point to depend on $J_0$ s only abolishing $J_1$ s. Note that
$J_0$ and $J_1$ has completely different behavior when the argument is small, the former is a decreasing function but the later is an
increasing function of the argument.

\begin{figure}
  \includegraphics[width=0.49\linewidth]{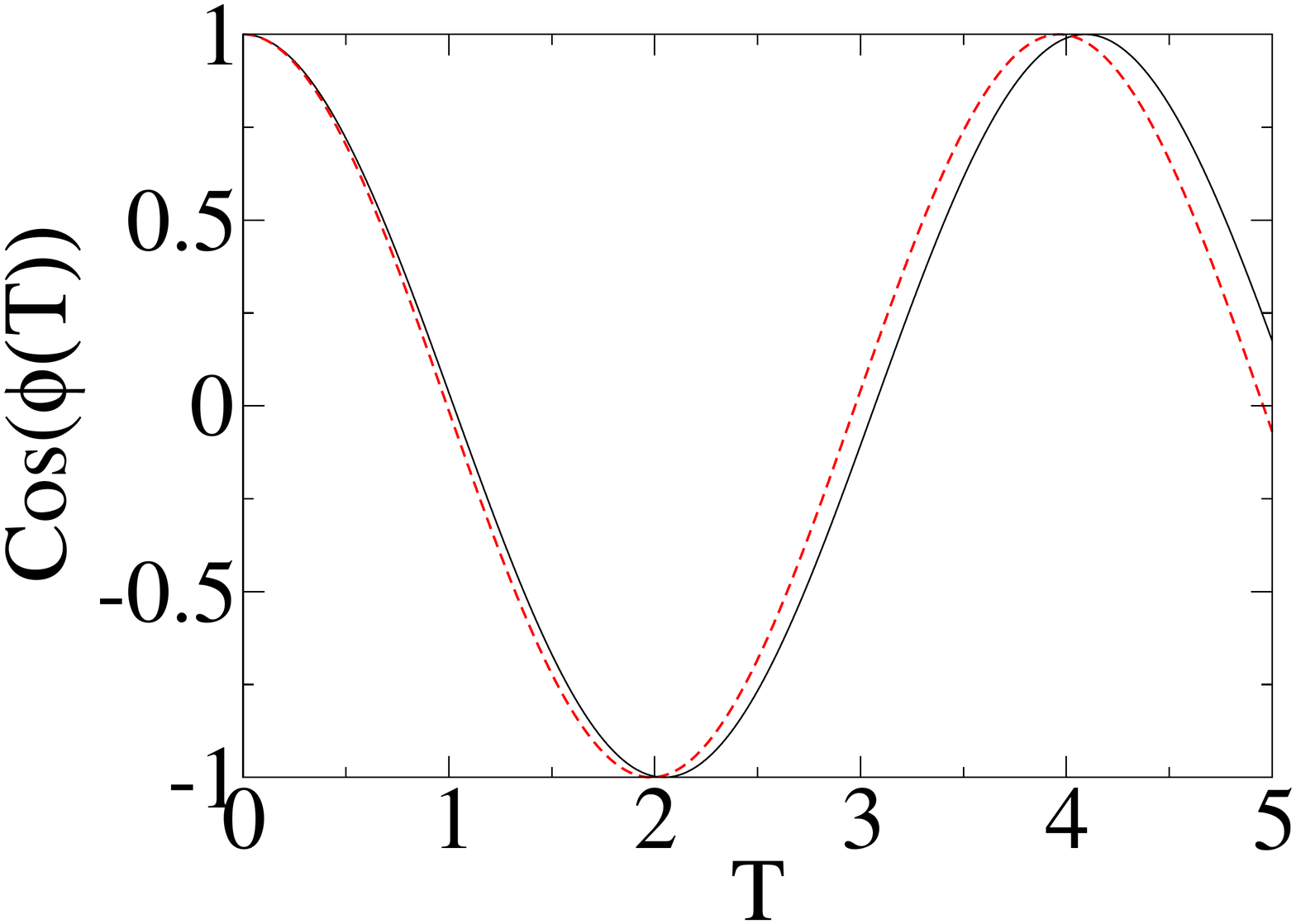}
\includegraphics[width=0.49\linewidth]{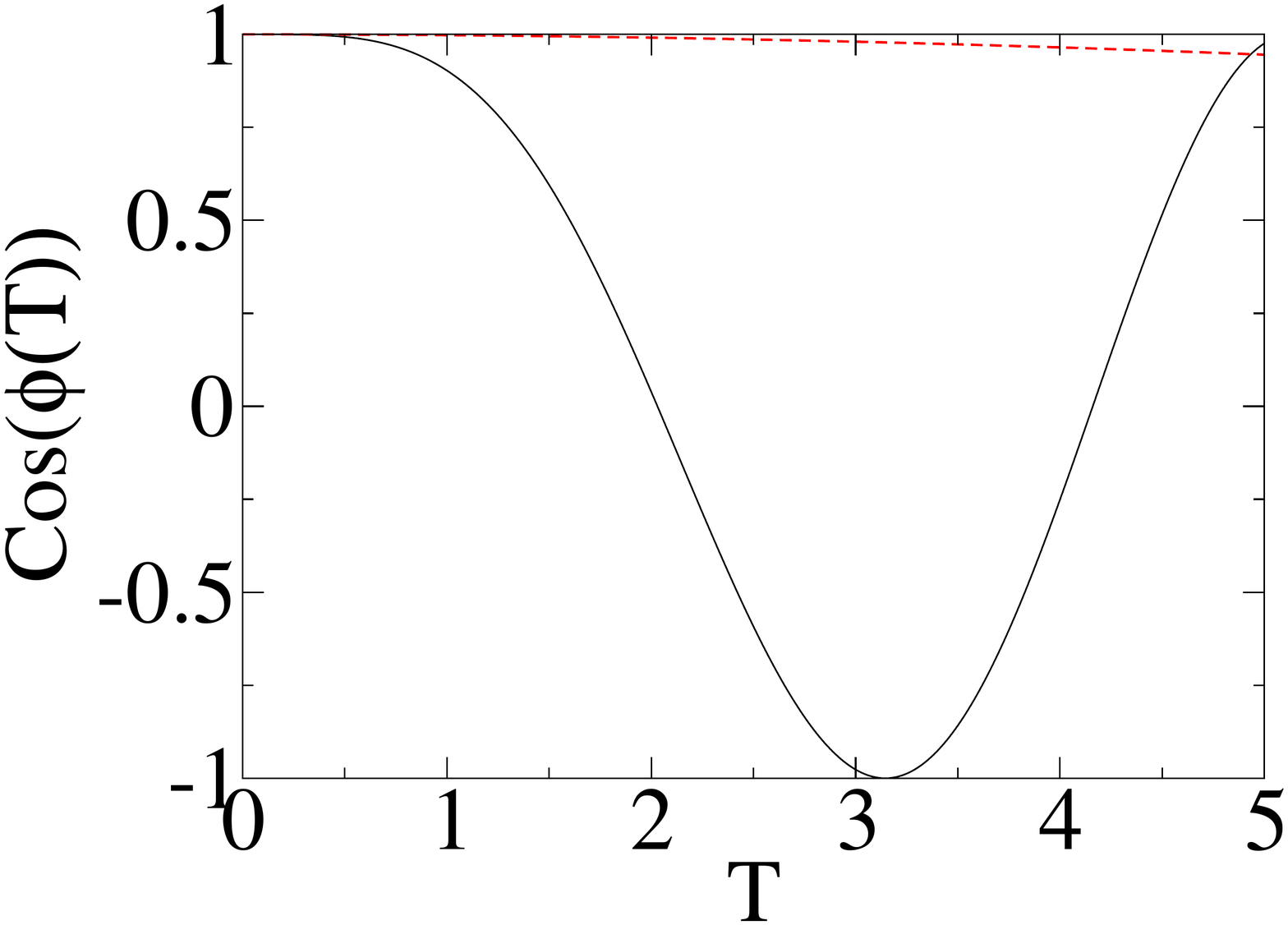}  
\caption{Comparison of cosines of Eq.\ref{(10)}(black-solid) and \ref{(14)}(red-dashed) for $\Gamma$ point (left panel) and of 
Eq.\ref{(11)}(black-solid) and \ref{(15)}(red-dashed)
for $K$ point (right panel). All parameters are same as before.}
\end{figure}

\subsection{Low frequency}\label{IIC}
At low frequencies (and also at high radiation amplitudes) one need to
take into account the higher Fourier components of the time-dependent Hamiltonian and consequently the truncation dimension
of the Floquet Hamiltonian increases. This is why at low frequencies one can't have simple analytical expression of Floquet bands 
in terms of Bessel functions and one needs to consider other methods like the adiabatic-impulse which gives good matching 
with numerics in low to moderate frequencies and high amplitudes\cite{us}. Symmetries of $H(t)$ also play a crucial role in predicting the existence
of phaseband crossings at different high symmetry points. But before going into the details of that we investigate the behavior of $D_N$ and $S_N$
as a function of N at low frequencies. Generally low $\omega$ and hence a high period ($T$) necessitates a proportional
increase of no of partitions but numerics suggests that the convergence of these quantities to zero is much slower than that in this parameter regime. In Fig.6(a)-(c) we
demonstrate this. We see for a typical high $\sigma$ one needs to increase N nearly quadratically (instead of linearly) with T to make the 
value of $D_N$ go below some particular threshold. We, therefore to reduce the numerical cost, keep our all calculations confined within small $\sigma$ values at low frequencies. 

\begin{figure}
\includegraphics[width=0.49\linewidth]{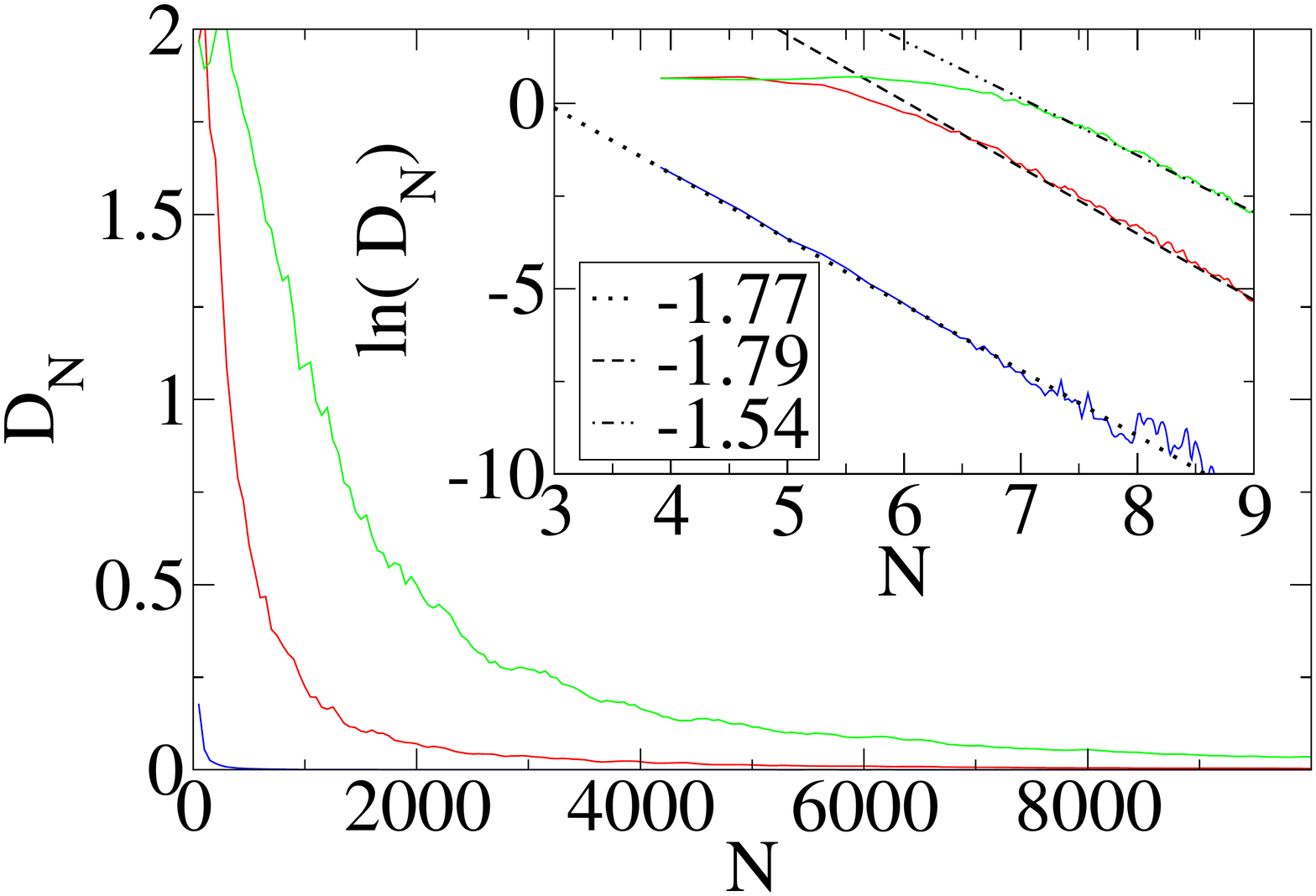}
\includegraphics[width=0.49\linewidth]{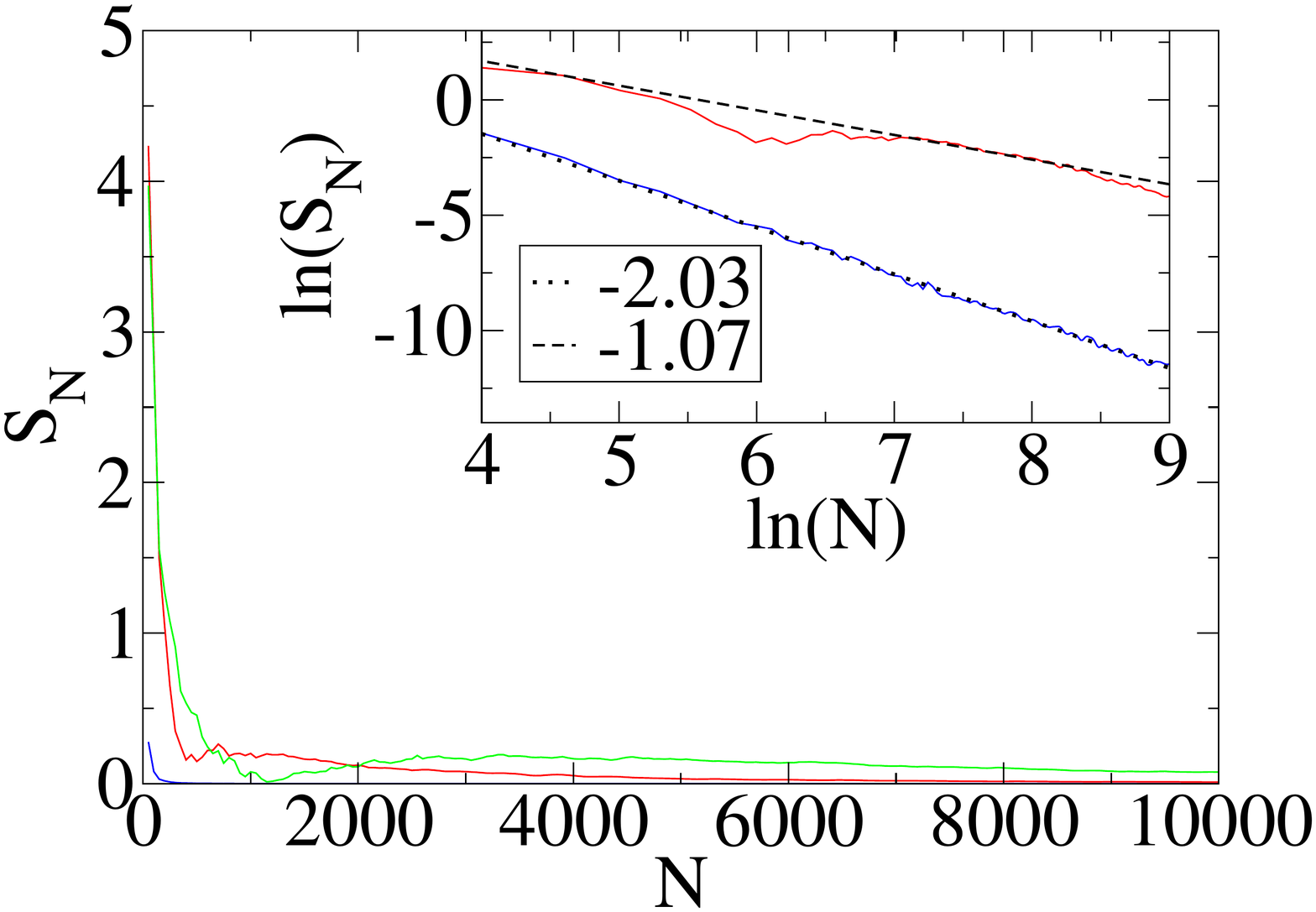} 
\includegraphics[width=0.49\linewidth]{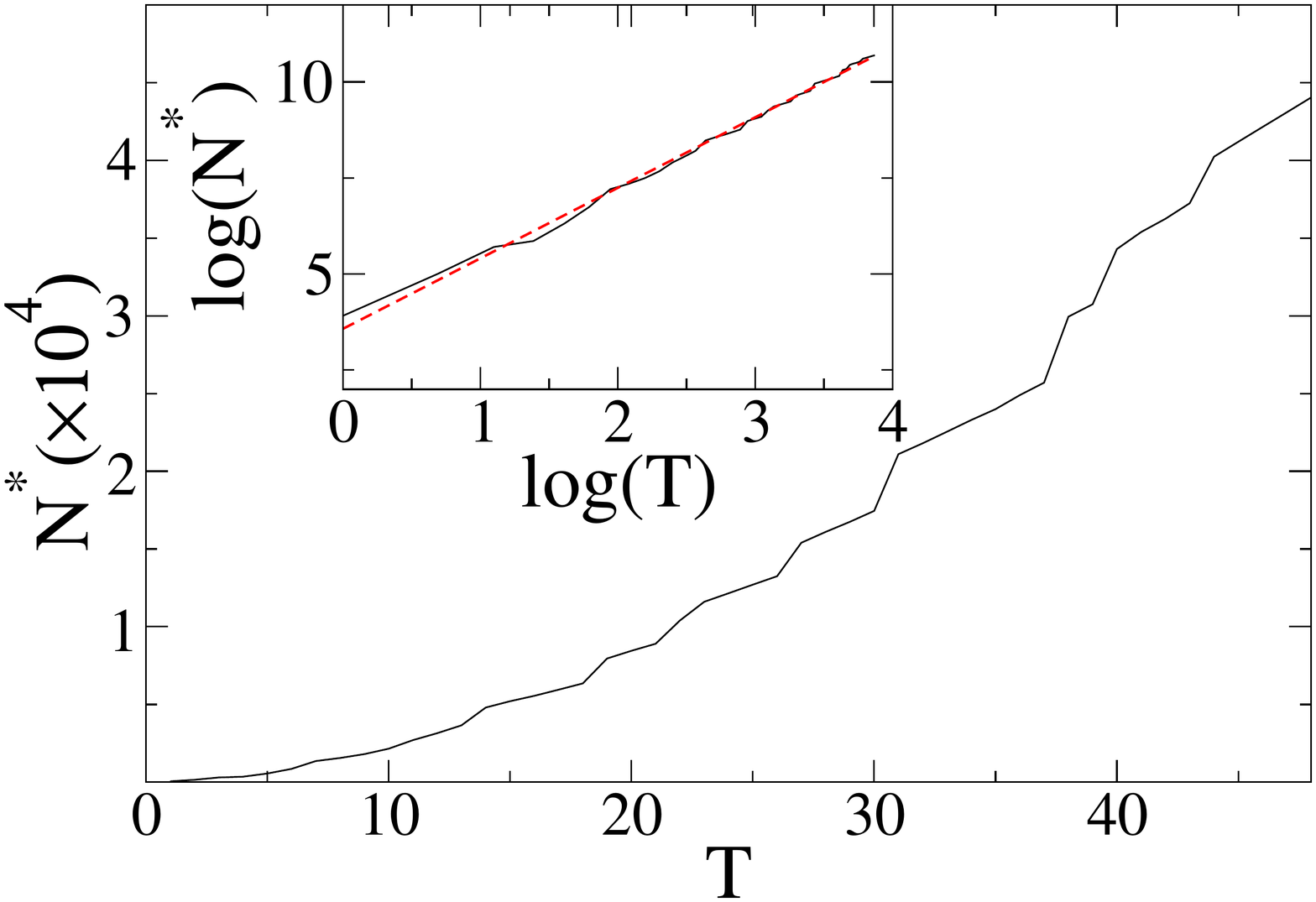} 
\includegraphics[width=0.49\linewidth]{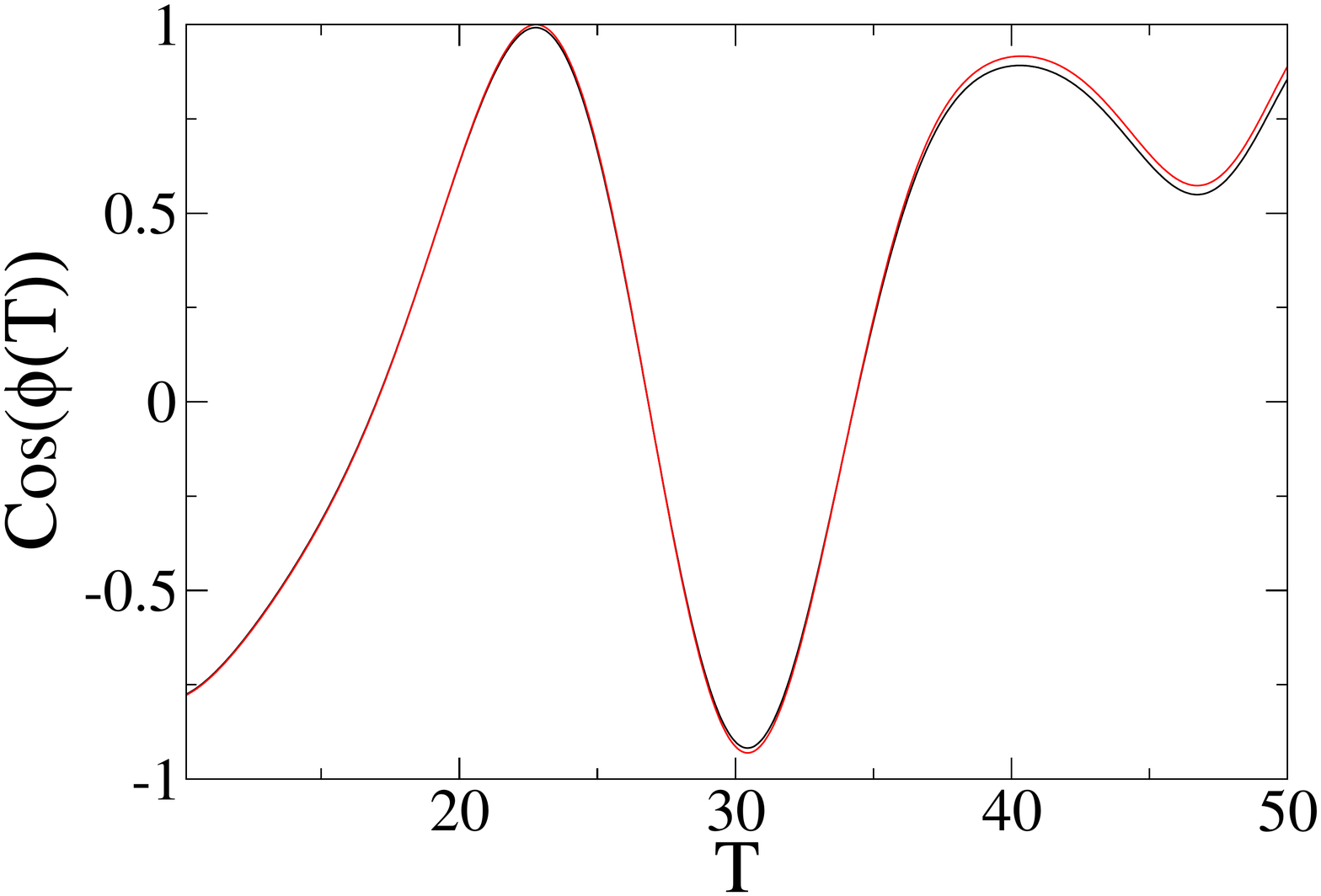} 
\caption{Fall of $D_N$(upper left panel) and $S_N$(upper right panel) with N for $\Gamma$ point at $\alpha=2.0$ and $T=60$. $\sigma$
for the blue,red and green curve is $\pi/50$, $\pi/10$ and $\pi/3$ respectively. $N^*$(for which $D_{N^*}$ fall below $10^{-4}$) vs T
in lower left panel for $\sigma=\pi/3$. Slope of linear fit in log-log plot is 1.83(inset). Lower right panel shows matching of phaseband from numerically averaged U(T) and U(T) calculated from averaged H
for $\Gamma$ point at $\alpha=2.2$ and $\sigma=\pi/10$.}
\end{figure}

It was shown in ref[33]\cite{us} that there exists 6 fold symmetries at $\Gamma$ point of graphene irradiated by circularly polarized light.
This was shown to be responsible for phaseband crossing simultaneously at $T/3$, $2T/3$ and $T$. But here for unpolarized
light typically all these symmetries are absent for any disorder-realization. Consequently, disorder averaging also leads to
avoided crossing. Here also the ensemble averaged Hamiltonian can capture the essential physics but interestingly
two of the symmetries get restored in it.  We chart out the 
symmetries of $\Gamma$ point under the irradiation of CP and unpolarized(ensemble averaged $H(t)$) light in detail in Table.1. This kind of symmetry mismatch between the two quantities inside the norm of Eq.\ref{(6)}
has significant impact on fall of $D_N$ at low frequencies. We find that $D_N$ falls very slowly with N (see Fig.6) here.

\begin{center}
\begin{table}
\begin{tabular}{|p{3.5cm}|p{1.5cm}|p{1.5cm}|}
\hline
\vskip 0.1cm \hskip 0.8cm Symmetries & \multicolumn{2}{|p{4cm}|}{\hskip 0.4cm Type of Polarization}\\ \cline{2-3}
           & \hskip 0.5cm CP & Unpolarized\\ 
\hline
$H(T-t)=H(t)$ & \hskip 0.6cm \color{green}\ding{52} & \hskip 0.8cm \color{green}\ding{52}\\
\hline
$H(\frac{T}{2}\pm t)=\tau_x H(t) \tau_x$ & \hskip 0.6cm \color{green}\ding{52} & \hskip 0.8cm \color{green}\ding{52}\\
\hline
$H(\frac{T}{6}\pm t)=\tau_x H(t) \tau_x$ & \hskip 0.6cm \color{green}\ding{52} & \hskip 0.8cm \color{red}\ding{56}\\
\hline
$H(\frac{T}{3}\pm t)=H(t)$ & \hskip 0.6cm \color{green}\ding{52} & \hskip 0.8cm \color{red}\ding{56}\\
\hline
$H(\frac{2T}{3}\pm t)=H(t)$ & \hskip 0.6cm \color{green}\ding{52} & \hskip 0.8cm \color{red}\ding{56}\\
\hline
$H(\frac{5T}{6}\pm t)=\tau_x H(t) \tau_x$ & \hskip 0.6cm \color{green}\ding{52} & \hskip 0.8cm \color{red}\ding{56}\\
\hline
\end{tabular}
\caption{Symmetries of $\Gamma$ point for circularly polarized and unpolarized light.}
\end{table}
\end{center}

In Fig.7 we show this symmetry mismatch between CP and unpolarized light pictorially (a large $\sigma$ is used for this purpose in Fig.7(a)) and its consequences. Fig.7(b)
shows for exact numerical disorder averaging a small $\sigma$ is sufficient to abolish the crossing at $T/3$. Fig.7(c)-(d) shows
for ensemble averaged Hamiltonian the crossings at $T/3$ and $2T/3$ gets increasingly avoided with increasing $\sigma$.
\begin{figure}
 \begin{center}
{\includegraphics[width=0.45 \columnwidth]{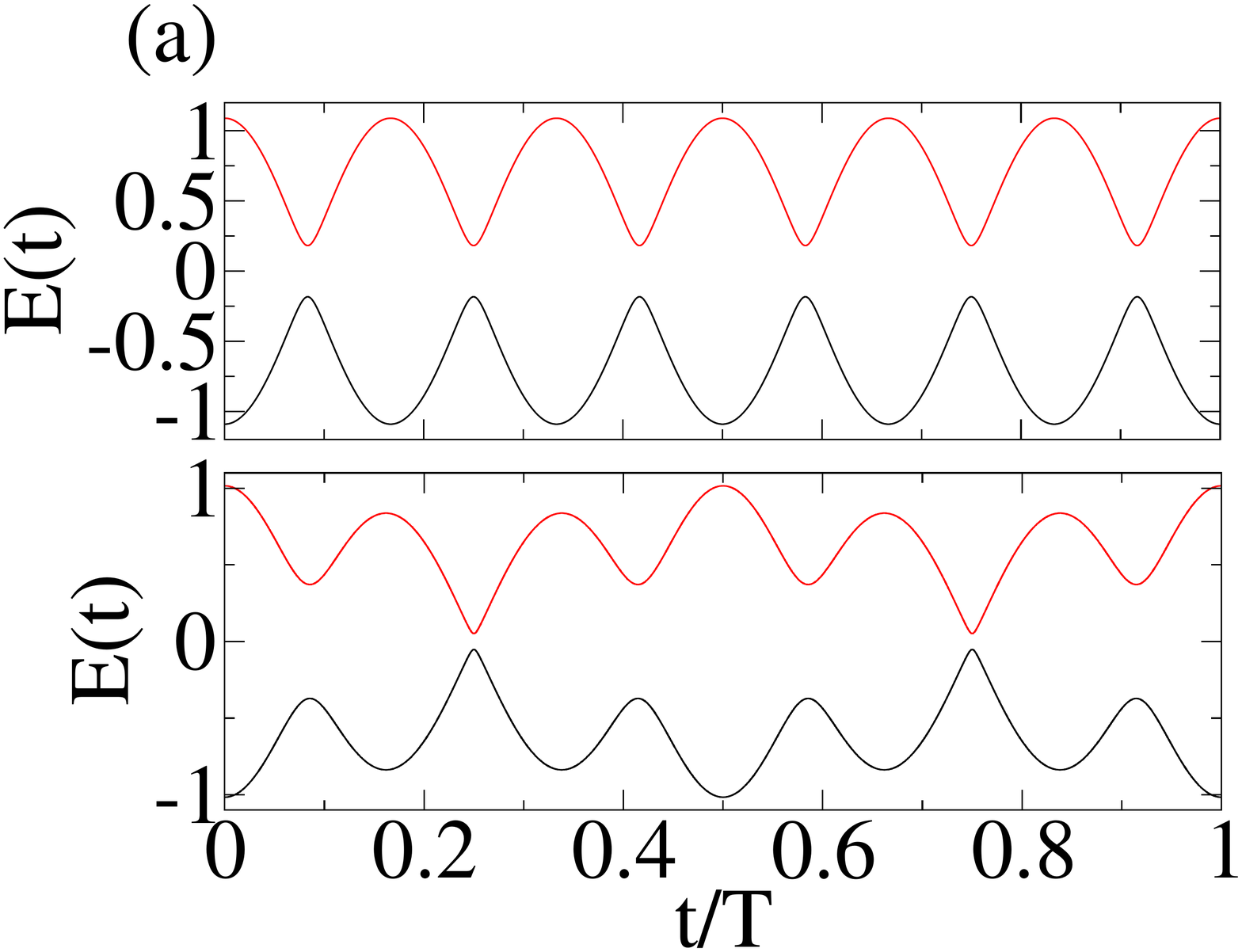}}
{\includegraphics[width=0.45 \columnwidth]{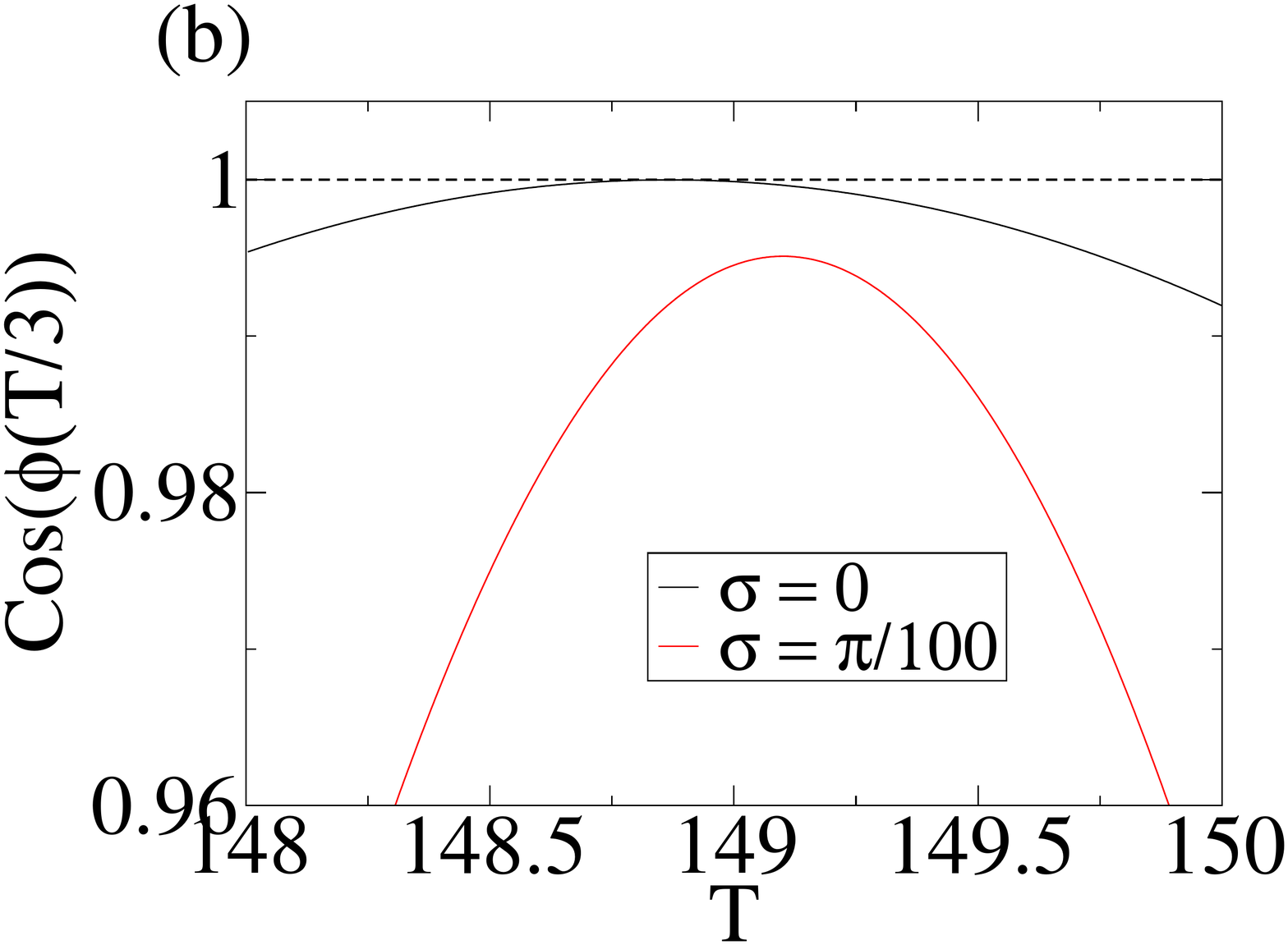}} \\
{\includegraphics[width=0.45 \columnwidth]{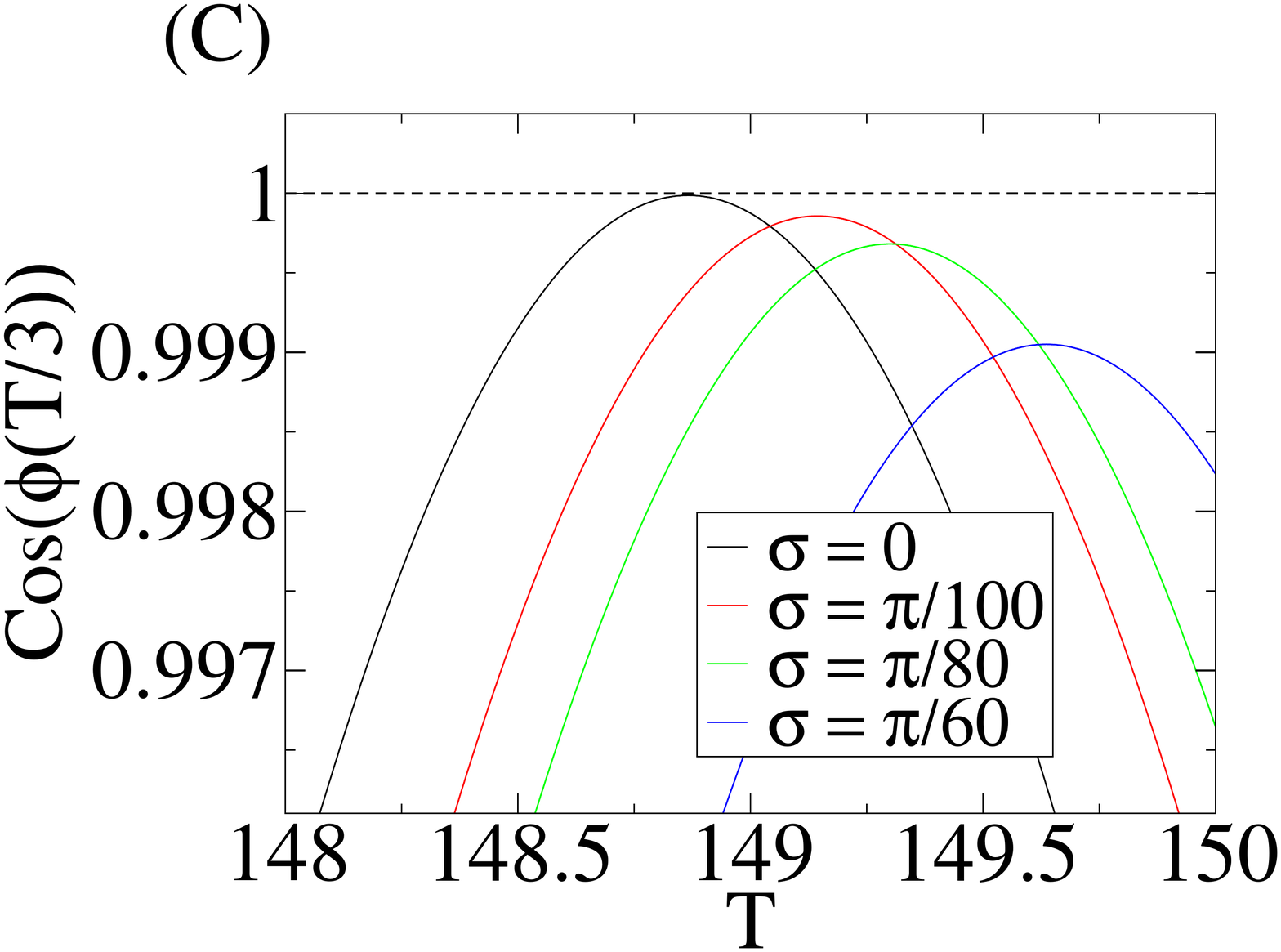}}
{\includegraphics[width=0.45 \columnwidth]{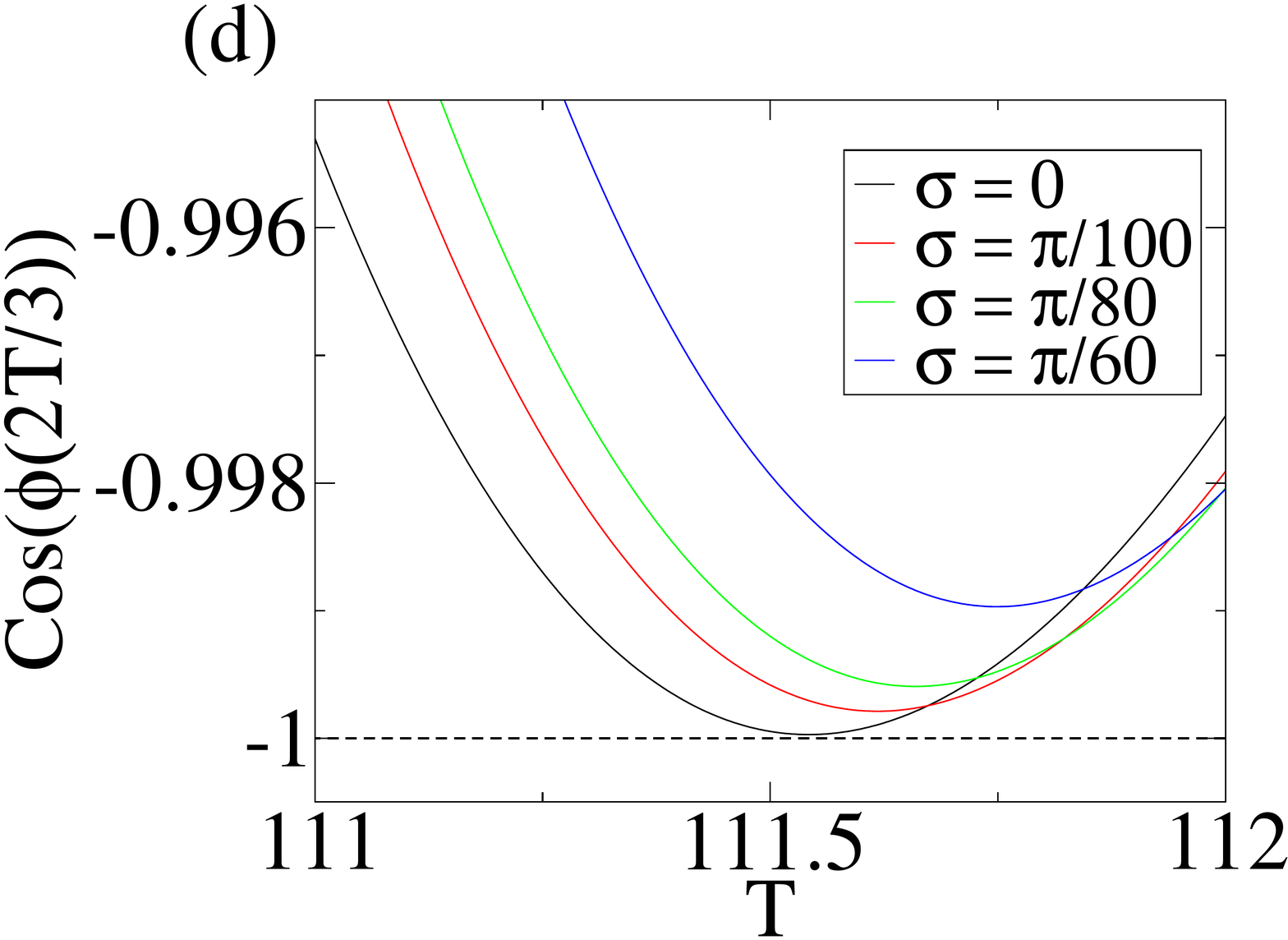}} \\
\end{center}
\caption{(a)Instantaneous energies vs t/T for CP(upper panel) and for unpolarized($\sigma=\pi/8$) in lower panel. $\alpha=2.3$
(b) $\cos(\Phi(\frac{T}{3}))$ vs T. The red curve is achieved by exact-numerical averaging with no of sample=1000, N=10000. $\alpha=2.35$ 
(c)$\cos(\Phi(\frac{T}{3}))$ vs T for $\alpha=2.35$ (d) $\cos(\Phi(\frac{2T}{3}))$ vs T for $\alpha=2.28$. Phasebands at (c)-(d) are
calculated using the ensemble averaged Hamiltonian}
\end{figure}
\section{1D systems}\label{III}
One-dimensional interacting spin chains whose Hamiltonian can be expressed in terms of free fermions via Jordan-Wigner transformation have attracted a lot
of theoretical attention in last decades due to their integrable structure, existence of topological transition as well as 
possibility of experimental realization using ion-traps and ultracold atom systems. Non-equilibrium dynamics in these models is equally interesting
because non-trivial topology can be induced by periodic drive of different terms in the Hamiltonian\cite{manisha}. This can be independently done
using multiple lasers with different amplitudes and frequency. In these experiments phase differences between different drive terms can be randomly changed
in a time scale $t_0\ll 1/\omega$ where $\omega$ is the frequency of drive. This constitutes a 1D platform to study similar 
physics as studied in previous section for 2D systems using unpolarized light. The survival of the topological transition under such
noisy drive is the key issue we would like to address. To this end, we consider a p-wave superconductor described by the following Hamiltonian\cite{amit1,amit2}
\begin{equation}\label{(16)}
 H=\sum_{i=1}^{L-1}[(\gamma c_i^\dagger c_i+H.c)+\Delta (c_ic_{i+1}+H.c)]-\mu\sum_{i=1}^L(2c_i^\dagger c_i-1)
\end{equation}
This model is equivalent to a spin-$\frac{1}{2}$ XY chain in perpendicular magnetic field via Jordan-Wigner transformation\cite{1961}. After a Fourier transformation defined by $c_k=\frac{1}{L}\sum_{j=1}^Lc_ie^{ikj}$ we can write this as
\begin{equation}\label{(17)}
 H=2\sum_{0\leq k \leq \pi}\psi_k^\dagger H_k \psi_k
\end{equation}
where $\psi_k=(c_k, c^\dagger_{-k})^T$ is a two component vector. Thus each k-mode of such systems can be described
by the following Hamiltonian(we scale everything by $\gamma$)
\begin{equation}\label{(18)}
 H(k,t)=(\mu-\cos(k))\sigma_z+\Delta \sin(k)\sigma_x
\end{equation}
and we use the following drive protocol $\mu=A\cos(\omega t+\phi(t))$ and $\Delta=\cos(r\omega t)$ where $r$ is an integer and 
$\phi$ is as usual a random variable at each time t. The 
dynamics of this model is non-trivial for $r>1$ due to the non-removable time dependence in both diagonal and off-diagonal element\cite{sau,satyaki}.
This model(with $\phi(t)=0$) has a phaseband crossing for $k=\pi/2$ at $t=T/2$ which exists at all frequencies. We study here what happens to this
crossing if at each instant of time $\phi$ is a random Gaussian variable with zero mean. Below we mention the scheme for partitioning
a full period to calculate the noise averaged $U(t,0)$ now at any time $t\leq T$
\begin{equation}\label{(19)}
 \delta t=\frac{t}{N}=const
\end{equation}
i.e we increase no of partitions proportionally as the time $t$ gets closer to $T$ keeping the duration of constant time evolution($\delta t$) fixed.
Thus we calculate noise averaged phaseband at all time t within a period for different noise strength ($\sigma$) and compare it with
noise free case in Figure.8(a). Interestingly noise modifies the phaseband at all times except at $t=T/2$ which is the 
phaseband crossing point for noise free drive. This shows that the transition at $t=T/2$ is immune to any amount of temporal disorder.
As a routine task we calculate the noise averaged instantaneous Hamiltonian for the chosen protocol
\begin{equation}\label{(20)}
 \langle H(k=\frac{\pi}{2},t) \rangle=A \cos(\omega t) e^{-\sigma^2/2}\sigma_z+\cos(r \omega t)\sigma_x
\end{equation}
\begin{figure}
 \includegraphics[width=0.49\linewidth]{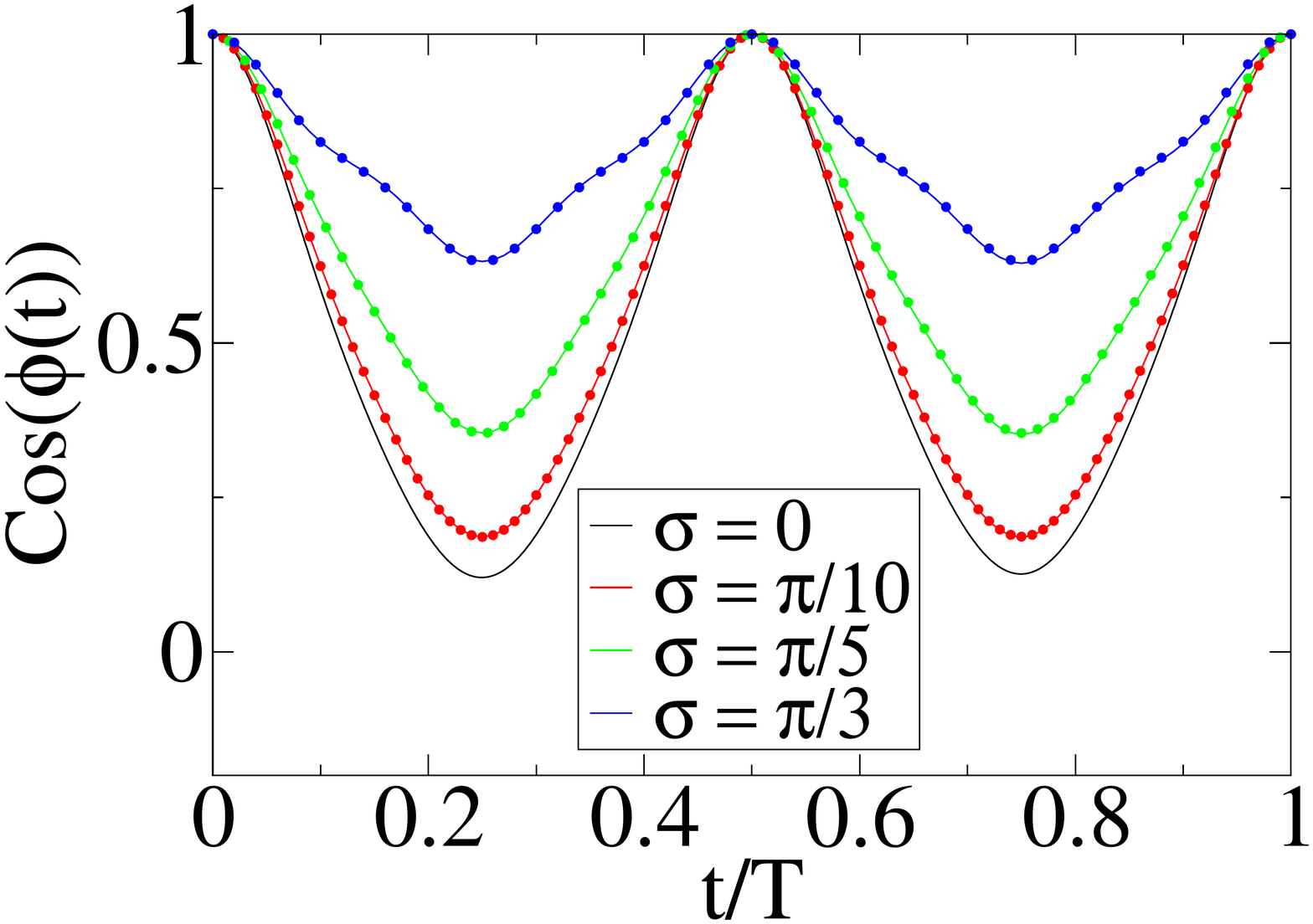}
\includegraphics[width=0.49\linewidth]{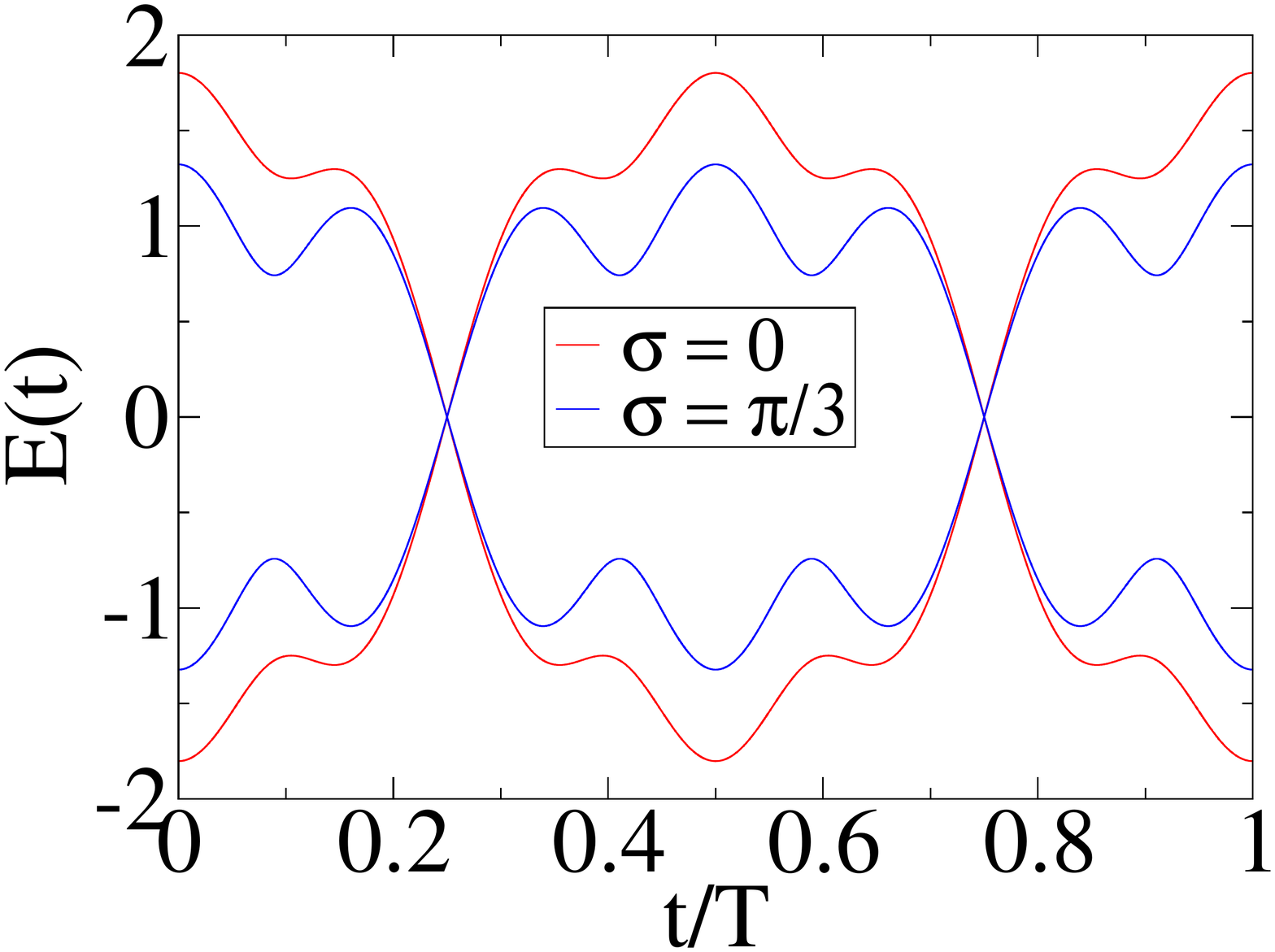} 
\caption{Phasebands from numerically averaged U operator (continuous line) and from the averaged Hamiltonians(dots) for 1D model(in Eq.\ref{(15)}) in left panel. 
A=1.5, $\omega=1.0$, r=3. Right panel shows the change of instantaneous energies with the insertion of noise.}
\end{figure}

In Fig.8(left panel) we see time evolution governed by this averaged $H$ mimics the numerically disorder averaged U operator as like before. 
We note that this numerical agreement leads to the following statement \textquotedblleft  The effect of random noise is just to renormalize
the laser amplitude \textquotedblright
\begin{equation}\label{(21)}
 \tilde{A}=Ae^{-\sigma^2/2}
\end{equation}
The robustness of the transition at $t=T/2$ also follows from the symmetry of Eq.\ref{(18)}. Note that the symmetry of the noise free Hamiltonian for $k=\pi/2$ and for odd $r$ 
(namely $H(T/2-t)=-H(t)$ )is not destroyed by the insertion of noise here (see Fig.8(right panel)). This can be used together with the Trotter like decomposition
of U operator (as in Eq.\ref{(2)}) to show $U^{-1}(T/2)=U^\dagger(T/2)=U(T/2)$ signifying that a crossing through Floquet zone-center will
always be there at $t=T/2$ for all parameter values ($A$, $\omega$, $\phi$ etc). Further right panel of Fig.8 demands that the same
adiabatic-impulse method (as done for the noise free case in ref.[33]) can be used to show the existence of the crossing at $t=T/2$ in spite of the change in sizes of different adiabatic
regions.
\section{Discussion}\label{IV}
In this work we have studied the existence of self-averaging limit in graphene irradiated by unpolarized light. 
We see the limit holds in high-frequency regime and can be captured by the noise-averaged Hamiltonian. In low frequencies the 
limit is achieved very slowly as a possible consequence of retaining two of the symmetries in noise-averaged Hamiltonian.
This opens up an opportunity to search for some  other deterministic Hamiltonian for speeding up the convergence to asymptotic limit. 
We hardly found any steady limit at extremely low frequencies to the best of our numerical ability. Floquet topological transitions are found to be modified by the insertion of noise to various degrees 
depending on the k-point in BZ. These
range from a small shift in crossing positions to complete abolition of the transition depending on the amount of disorder. We
find that certain k-points are more affected as a consequence of a change in Fourier structure of their time-dependent 
Hamiltonian 
induced by the noise. The presence of a 6-fold symmetry at $\Gamma$ point plays a crucial role for the existence of a special type of
crossings which simultaneously happens at $T/3$,$2T/3$,$T$\cite{us}. This kind of crossings are ubiquitous in low frequencies but
ceases to exist in high frequency(scanning the whole parameter regime as much as possible we found they are absent below $T\approx11$). Now breaking of
4 out of those 6-symmetries by the noise abolishes these transitions confirming again the importance of symmetries in low frequencies. 
In 1D systems due to the simplicity of the BZ, noise obeys all symmetries of the clean time-dependent Hamiltonian and as a consequence crossings persist at all 
noise strengths. It merely renormalizes the drive amplitude.

In typical experiments one needs to keep the optical axis of a quarter wave plate exactly at $45{^\circ}$ with the plane of 
vibration of the incident plane polarized light to extract pure circularly polarized light. Now if this angle changes randomly (which is always present in small amount if the experiment
is not performed carefully such as a small vibration of the table on which the set up lies may cause it) then the polarization of the 
outgoing light will also fluctuate. One can also use synthetic gauge fields to produce such noisy vector potential. This kind of perturbation is very common
in an interference experiment if incoherent sources are used. 
The quantitatively different noise-response from various k-points can be experimentally verified by measuring the photoinduced gap in a
momentum resolved manner using pump-probe spectroscopy as done in ref[22]. The abolition of transition and hence a change in topological structure 
of the Floquet bands can be detected by analyzing the intensity and angular dependence of ARPES spectra\cite{gavensky}.

In conclusion we have shown random noise in the vector potential of incident light has significant impact on Floquet topological
transition in graphene. One can analyze the symmetries and Fourier structure of the noise-averaged Hamiltonian to understand 
the modifications done by the noise. In 1D systems such noisy drives has no effect on the transitions.

\section*{Acknowledgements}
Author thanks K. Sengupta for support.

\end{document}